# von Neumann Stability Analysis of Globally Constraint-Preserving DGTD and PNPM Schemes for the Maxwell Equations using Multidimensional Riemann Solvers

By

**Dinshaw S. Balsara[1] and Roger Käppeli[2]**

[1]University of Notre Dame (dbalsara@nd.edu)

[2]Seminar for Applied Mathematics (SAM), Department of Mathematics, ETH Zürich, CH-8092 Zürich, Switzerland (roger.kaeppeli@sam.math.ethz.ch)

**Abstract**

The time-dependent equations of computational electrodynamics (CED) are evolved consistent with the divergence constraints on the electric displacement and magnetic induction vector fields. Respecting these constraints has proved to be very useful in the classic finite-difference time-domain (FDTD) schemes. As a result, there has been a recent effort to design finite volume time domain (FVTD) and discontinuous Galerkin time domain (DGTD) schemes that satisfy the same constraints and, nevertheless, draw on recent advances in higher order Godunov methods. This paper catalogues the first step in the design of globally constraint-preserving DGTD schemes. The algorithms presented here are based on a novel DG-like method that is applied to a Yee-type staggering of the electromagnetic field variables in the faces of the mesh. The other two novel building blocks of the method include constraint-preserving reconstruction of the electromagnetic fields and multidimensional Riemann solvers; both of which have been developed in recent years by the first author.

The resulting DGTD scheme is linear, at least when limiters are not applied to the DG scheme. As a result, it is possible to carry out a von Neumann stability analysis of the entire suite of DGTD schemes for CED at orders of accuracy ranging from second to fourth. The analysis requires some simplifications in order to make it analytically tractable, however, it proves to be extremely instructive. A von Neumann stability analysis is a necessary precursor to the design of a full DGTD scheme for CED. It gives us the maximal CFL numbers that can be sustained by the DGTD schemes presented here at all orders. It also enables us to understand the wave propagation characteristics of the schemes in various directions on a Cartesian mesh. We find that constraint-preserving DGTD schemes permit CFL numbers that are competitive with conventional DG schemes. However, like conventional DG schemes, the CFL of DGTD schemes decreases with increasing order. To counteract that, we also present constraint-preserving PNPM schemes for CED. We find that the third and fourth order constraint-preserving DGTD and P1PM schemes have some extremely attractive properties when it comes to low-dispersion, low-dissipation propagation of electromagnetic waves in multidimensions. Numerical accuracy tests are also provided to support the von Neumann stability analysis. We expect these methods to play a role in those problems of engineering CED where exceptional precision must be achieved at any cost.

## I) Introduction

The numerical solution of Maxwell's equations plays a crucial role in the treatment of many problems in science and engineering. The finite-difference time-domain (FDTD) method (Yee [47], Taflove [43], Taflove and Hagness [44], [46], Taflove, Oskooi and Johnson [45]) has been a primary technique for this class of computational electrodynamics (CED) applications for more than a quarter century. The staggering of variables in the FDTD method gives it many desirable features, including a direct interpretation of the two curl-type equations given by Faraday's Law and the generalized Ampere's Law, and a natural satisfaction of the constraint equations given by Gauss's Laws for electric and magnetic charge. On a simple Cartesian mesh, every electric field vector component is surrounded by four circulating magnetic field vector components, and every magnetic field vector component is surrounded by four circulating electric field vector components. This compactly staggered arrangement of primal variables is the source of the FDTD method's strength and versatility. Realize though that the FDTD scheme is built on a philosophy that predates many advances in the numerical treatment of hyperbolic systems.

We realize that Maxwell's equations are a hyperbolic system, and the last few decades have seen dramatic advances in the numerical solution of hyperbolic systems. These general-purpose methods go under the name of higher order Godunov schemes and in the last decade or so they have also been developed to a high degree of sophistication. The design philosophy underlying higher order Godunov schemes is very general and applies to any hyperbolic system. Such schemes enable us to analyze each of the individual waves that propagate in a hyperbolic system and provide a sophisticated higher order treatment of the system as a whole. High conductivity at conductors can also make the hyperbolic system stiff, and higher order Godunov schemes offer some very elegant A-stable ways of treating those stiff terms. Formulations that treat Maxwell's equations with zone-centered higher order Godunov methods have been tried (Munz *et al*. [36], Ismagilov [33], Barbas and Velarde [22]; and references therein). Because these methods are based on finite volume approaches, they are often referred to as finite volume time domain (FVTD) methods. However, in their native form, these FVTD methods for CED do not have the ability to preserve the mimetic constraints inherent in Faraday's Law and the generalized Ampere's Law. Variants of higher order Godunov schemes have been developed for treating the magnetohydrodynamic (MHD) equations which do indeed respect the mimetic constraints in Faraday's Law. A constraint-preserving reconstruction strategy was crucial to making this advance (Balsara [2], [3], [4], Balsara and Dumbser [12], Xu *et al*. [48], Balsara *et al*. [16]). The development of multidimensional Riemann solvers (Balsara [7], [8], [11], [14], Balsara, Dumbser and Abgrall [10], Balsara and Dumbser [13], Balsara *et al*. [15], Balsara and Nkonga [21]) was another crucial step in making this breakthrough. In a sequence of recent papers (Balsara *et al*. [16], [19], [20]) these advances have also been extended to design constraint-preserving, higher order Godunov, FVTD schemes for CED.

The Discontinuous Galerkin (DG) method is unique amongst variants of higher order Godunov methods. The method was first developed by Cockburn & Shu [24], [25], Cockburn, Hou & Shu [26] based on an initial proposal by Reed and Hill [37]. Except for the mean value, all the modes in the FVDT schemes described in the previous paragraph are reconstructed at the start of each new timestep. DG methods are different in the sense that all the modes are evolved in time with the help of the governing equations. This is done by asserting that within each zone we have a set of trial functions onto which the solution is projected. The modes are then evolved by making a Galerkin projection of the governing equation with the help of a set of test functions. Typically, the test functions are chosen from the same space as the trial functions. In some sense, the DG method is analogous to a spectral scheme in that it relies on a Galerkin projection of a complete basis space. This accounts for the method's very high accuracy and fidelity. For linear equations, like Maxwell's equations, the method becomes even closer to a spectral scheme; and this spectral-like accuracy makes it very attractive for high precision engineering CED calculations. The spectral-like accuracy highlights the value of developing DG schemes for CED. However, previous generations of DG schemes for CED were not globally constraint-preserving (Hesthaven and Warburton [32], Cockburn, Li and Shu [28] Kretzschmar *et al.* [34], Egger *et al.* [30], Bokil *et al.* [23]). This prevents a harmonious blending of the best attributes of DGTD schemes with the best attributes of FDTD schemes. DG schemes for CED are sometimes referred to as DGTD (discontinuous Galerkin time domain) schemes and we continue that notation in this paper.

In their study of the induction equation for MHD, Balsara & Käppeli [18] devised a globally constraint-preserving DG scheme that exactly preserved the constraints inherent in Faraday's law. However, that scheme was strongly oriented towards the MHD equations. Nevertheless, the essential ingredients of a globally constraint-preserving DG scheme for CED are clearly anticipated in that work. Their algorithm, therefore, relied on endowing higher moments to the components of the magnetic induction that reside in the faces of the mesh. The facial modes were, therefore, the primal variables of the scheme. From these facial moments, a constraint-preserving reconstruction could be performed that made the magnetic field available at all locations on the mesh (please see Fig. 1 of Balsara & Käppeli [18]). A Galerkin projection applied to each face of the skeleton mesh then provided the evolutionary equations for the facial modes. A finite volume DG scheme relies on a weak form representation of the Riemann solver-based numerical flux at each zone face. In an exactly analogous fashion, the constraint-preserving DG scheme for the induction equation had to rely on a weak form representation of the electric field at each edge of the mesh. This is where Balsara & Käppeli [18] utilized a multidimensional Riemann solver for the induction equation. In Balsara *et al.* [16], [19] a multidimensional Riemann solver was designed for CED. This innovation opens the door to globally constraint-preserving DGTD schemes for CED.

Before a full multidimensional, globally constraint-preserving DGTD scheme can be designed for CED, one has to carry out a von Neumann stability analysis of the said scheme. Please see Fig. 4.2 of Taflove and Hagness [44] to get very important insights into the dispersion properties of the FDTD scheme. Such a von Neumann stability analysis is necessarily simpler than the full scheme, ignoring the role of material media and stiff source terms. Despite these simplifications, such a von Neumann stability analysis for FDTD has proved to be extremely instructive. In fact, a von Neumann stability analysis is an essential precursor to the design of a full scheme because it would give us two very important insights. First, it would give us the acceptable range of CFL numbers for the DGTD schemes. Second, it would give us insight on the wave propagation characteristics of the DGTD schemes. An analogous von Neumann stability analysis for multidimensional, globally constraint-preserving DGTD schemes for CED has never been carried out. Indeed, before the advent of the multidimensional Riemann solver for CED in Balsara *et al*. [16], [19], the formulation of such constraint-preserving DGTD schemes that use the same control volume was out of question. In this paper, we formulate a simplified, two-dimensional, globally constraint-preserving DGTD scheme for CED in homogeneous and isotropic non-conductive media and carry out the von Neumann stability analysis for the scheme. In a subsequent paper we will present a full three-dimensional globally constraint-preserving DGTD scheme for CED that can handle material media with strong variations in permittivity, permeability and conductivity.

For conventional DG schemes, as well as the DGTD schemes designed here, the permissible CFL decreases with increasing order of accuracy. This trend diminishes the practical utility of DGTD schemes. PNPM schemes (Dumbser *et al*. [29]) are one way of overcoming this problem. (PNPM schemes evolve an $N^{th}$ order spatial polynomial, while spatially reconstructing higher order terms up to $M^{th}$ order.) In Dumbser *et al*. [29] we found that P1PN or P2PN schemes retain most of the accuracy of a DG scheme while offering a much larger limiting CFL. In Balsara and Käppeli [18] we found that DG schemes for the induction equation show the same advantageous trend. For that reason, we design constraint-preserving PNPM schemes for CED in this paper and show that they have the same twin advantages of a larger limiting CFL and accuracies that rival those of constraint-preserving DGTD schemes for CED.

In light of the above discussion, we set four goals for this paper. Our *first goal* is to design a DGTD scheme for Maxwell's equations that is globally constraint-preserving. This goal is achieved with the help of two advances – globally constraint-preserving reconstruction and multidimensional Riemann solvers. The resulting von Neumann stability analysis can be very useful for studying the wave-propagation characteristics of DGTD schemes at various orders and it can also be useful for identifying the maximal CFL number that can be achieved by various schemes. Thus our *second goal* is to identify the wave propagation characteristics at various angles to a Cartesian mesh. We do this for DGTD schemes that are second, third and fourth order accurate. For each of these schemes, we are in a position to specify the dissipation and dispersion of waves with various wavelengths relative to the mesh size. We even compare our results to the results from Fig. 4.2 of Taflove and Hagness [44]. Our *third goal* is to identify the maximal CFL number associated with schemes with different orders of spatial accuracy that are updated with Runge-Kutta timestepping schemes with various orders of temporal accuracy. Many of our Runge-Kutta timestepping schemes are strong stability preserving (Shu and Osher [40], [41], Spiteri and Ruuth [38], [39]). Our study of the maximal CFL number for DG schemes follows the style of prior studies (Zhang and Shu [50], Liu et al. [35], Yang and Li [49], Balsara and Käppeli [18]). Our *fourth goal* is to show that constraint-preserving PNPM schemes for CED (with low values of "N") retain a large maximal CFL number while furnishing most of the inherent accuracy of DGTD schemes.

Section II presents the Maxwell's equations in a simplified format that is suitable for von Neumann stability analysis. Section III presents a second order, P=1, DGTD scheme for Maxwell's equations and explicits the von Neumann stability analysis for that case. Section IV presents some details for a third order, P=2, DGTD scheme for Maxwell's equations. Section V presents the results from the von Neumann stability analysis. Section VI presents numerical results. Section VII draws conclusions. Appendix A provides an explicit specification of the amplification matrix for the P=1 DGTD scheme. The equations for an analogous fourth order DGTD scheme are presented in an Appendix B.

**II) Simplified Structure of the Maxwell Equations; Suitable for the von Neumann Stability Analysis**

As in Taflove and Hagness [48], von Neumann analysis for CED is always carried out on a simplified equation set associated with the Maxwell's equations. Thus, we assume that the permittivity and permeability are simple scalars with no spatial variation. Consequently, the permittivity $\varepsilon$ and the permeability $\mu$ are taken to be constants. We wish to use the von Neumann stability analysis to understand wave propagation in CED. As a result, the current densities and charge densities are set to zero. From Faraday's law for the evolution of the magnetic induction we get

$$\frac{\partial \mathbf{B}}{\partial t} + \frac{1}{\varepsilon} \nabla \times \mathbf{D} = 0 \tag{2.1}$$

From the extended Ampere's law for the evolution of the electric displacement we get

$$\frac{\partial \mathbf{D}}{\partial t} - \frac{1}{\mu} \nabla \times \mathbf{B} = 0 \tag{2.2}$$

The vector fields satisfy involution constraints. The equations of constraint are also given by

$$\nabla \cdot \mathbf{B} = 0 \tag{2.3}$$

and

$$\nabla \cdot \mathbf{D} = 0 \tag{2.4}$$

Because of their vector nature, Maxwell's equations should always be treated three-dimensionally. To simplify the stability analysis, we assume that all significant variations are restricted to the two-dimensional *xy*-plane. The z-component of $\mathbf{D}$ is assumed zero and the *x*- and *y*-components of $\mathbf{B}$ are also assumed zero. In other words, we focus on a $TE_z$ mode. (Our formulation, which is based on the multidimensional Riemann solver, is entirely symmetrical and we could as well have analyzed the $TM_z$ mode without any change to the analysis. Realize that with the transcriptions $\mathbf{B} \to \mathbf{D}$, $\mathbf{D} \to \mathbf{B}$, $\varepsilon \to -\mu$ and $\mu \to -\varepsilon$ we see that a $TE_z$ mode becomes a $TM_z$ mode.) The evolutionary equations can be written in flux form as

$$\frac{\partial}{\partial t}\begin{pmatrix} D_x \\ D_y \\ B_z \end{pmatrix} + \frac{\partial}{\partial x}\begin{pmatrix} 0 \\ B_z/\mu \\ D_y/\varepsilon \end{pmatrix} + \frac{\partial}{\partial y}\begin{pmatrix} -B_z/\mu \\ 0 \\ -D_x/\varepsilon \end{pmatrix} = 0 \tag{2.5}$$

This is the equation set for which we will develop a von Neumann stability analysis at second order in Section III and at third order in Section IV. The Appendix details the fourth order case.

## III) Second Order, P=1, DGTD Scheme for Maxwell's Equations

In Sub-section III.a we formulate the second order, P=1, DGTD scheme for Maxwell's equations. In Sub-section III.b we describe its von Neumann stability analysis.

### III.a) Formulation of the Second Order, P=1, DGTD Scheme for Maxwell's Equations

We begin by pointing out that the method presented here is not a conventional DG scheme for conservation laws. Since it shares several features with a conventional DG scheme, which is why we call it a DG-like scheme. Section II of Balsara & Käppeli [18] provides a precise compare and contrast between the DG-like schemes that we formulate here and a conventional DG scheme. A conventional DG scheme for conservation laws is based on the Gauss law-based vector identity

$$\nabla \cdot (\phi \, \mathbf{F}) = \phi \, \nabla \cdot \mathbf{F} + \mathbf{F} \cdot \nabla \phi$$

By contrast, the present schemes live in the faces of a three-dimensional mesh because Faraday's law and the generalized Ampere's law are based on a Stokes law update. As a result, our DG-like scheme for updating curl-type equations is based on the Stokes law-based vector identity

$$\nabla \times (\phi \, \mathbf{D}) = (\nabla \phi) \times \mathbf{D} + \phi \, \nabla \times \mathbf{D} \tag{3.1}$$

We will assert this vector identity in the faces of a three-dimensional mesh.

Even though we are interested in a two-dimensional scheme, it is most beneficial to realize that eqns. (2.1) and (2.2) are actually three dimensional and to consider the DG formulation on a three-dimensional mesh. Specifically, let us first focus on eqn. (2.2). One zone of a three-dimensional mesh is shown in Fig. 1. The mesh is Cartesian and has uniform zones with size $\Delta x$, $\Delta y$ and $\Delta z$ in the $x$-, $y$- and $z$-directions. To keep things simple, the zone in Fig. 1 has an extent $[-\Delta x/2, \Delta x/2] \times [-\Delta y/2, \Delta y/2] \times [-\Delta z/2, \Delta z/2]$, although we will find that $\Delta z$ does not play a role in the update equations that we will finally derive. However, since we are most interested in a two-dimensional scheme, we will assume that all the spatial variation is restricted to having variations only in the $x$-direction or in the $y$-direction. We also restrict focus to TE$_z$ modes. Let $\hat{\mathbf{n}}$ be the unit outward pointing normal to an element of this mesh. Let $A_n$ be the area of the face to which $\hat{\mathbf{n}}$ is the unit normal. Say we take $\hat{\mathbf{n}} = \hat{\mathbf{x}}$ in Fig. 1; in that case the area $A_n$ will be the right

*x*-face which contains the field component $D_x$ that we are interested in evolving via eqn. (2.5). Now say we take $\hat{\mathbf{n}} = \hat{\mathbf{y}}$ in Fig. 1; in that case the area $A_n$ will be the upper *y*-face which contains the field component $D_y$ that we are interested in evolving via eqn. (2.5). We wish to project eqn. (2.2) into a space of test functions. Our test functions will be taken to be identical to our trial functions and we will make these trial functions explicit in a little while. To make the DG-like projection, we first multiply eqn. (2.2) with the test function $\phi$. Next, we restrict attention to the face $A_n$ by taking a dot product with the unit normal $\hat{\mathbf{n}}$ to that face. We then integrate over that face and use the vector identity in eqn. (3.1) to get

$$\frac{\partial}{\partial t}\left(\int_{A_n} (\hat{\mathbf{n}}\cdot\mathbf{D})\phi\, dA_n\right) - \frac{1}{\mu}\int_{\partial A_n}(\phi\,\mathbf{B})\cdot d\vec{\ell} + \frac{1}{\mu}\int_{A_n}\hat{\mathbf{n}}\cdot\left[(\nabla\phi)\times\mathbf{B}\right]dA_n = 0 \qquad (3.2)$$

The boundary of the face under consideration is denoted by $\partial A_n$. The infinitesimal vector $d\vec{\ell}$ in the middle term of eqn. (3.2) runs along $\partial A_n$ and denotes the length of the element. The existence of a unit normal, $\hat{\mathbf{n}}$, lends a right-handed directionality to $d\vec{\ell}$. Eqn. (3.2) gives us the desired Galerkin projection strategy; but please realize that it applied to a curl-type equation in the faces of the mesh. Notice that the second term in eqn. (3.2) is interpreted in a weak form using a multidimensional Riemann solver and is analogous to the flux term in a traditional DG method for conservation laws. The third term in eqn. (3.2) is analogous to the volume term in a traditional DG method for conservation laws. Eqn. (3.2) is our master equation that stems from the extended Ampere's law. In the next few paragraphs we will show how it is to be used to design a DG scheme at second order.

Let us now instantiate eqn. (3.2). In the right *x*-face of the element shown in Fig. 1, i.e. the face with $x = \Delta x/2$, we assert the second order accurate evolution of the *x*-component of the displacement vector to be of the form

$$D^x(y,t) = D_0^x(t) + D_y^x(t)\left(\frac{y}{\Delta y}\right) \qquad (3.3)$$

The variation of the *x*-component of the electric displacement in eqn. (3.3) is also shown in Fig. 1. Notice that we have suppressed the *z*-variation in the above equation because we are only interested in a two-dimensional scheme. Our trial functions are $\phi(y)=1$ and $\phi(y)=(y/\Delta y)$. Using $\hat{\mathbf{n}} = \hat{\mathbf{x}}$ and the test function $\phi(y)=1$, eqn. (3.2) then gives us

$$\frac{dD_0^x(t)}{dt} - \frac{1}{\mu}\frac{1}{\Delta y}\left[B^{z**}(x=\Delta x/2, y=\Delta y/2) - B^{z**}(x=\Delta x/2, y=-\Delta y/2)\right] = 0 \tag{3.4}$$

Using $\hat{\mathbf{n}} = \hat{\mathbf{x}}$ and the test function $\phi(y)=(y/\Delta y)$ in eqn. (3.2) we get

$$\frac{1}{12}\frac{dD_y^x(t)}{dt} - \frac{1}{\mu}\frac{1}{2\Delta y}\left[B^{z**}(x=\Delta x/2, y=\Delta y/2) + B^{z**}(x=\Delta x/2, y=-\Delta y/2)\right]$$
$$+ \frac{1}{\mu}\frac{1}{\Delta y}\left\langle B^{z*}(x=\Delta x/2, y)\right\rangle = 0 \tag{3.5}$$

Eqn. (3.2) is crucially important for deriving the above two equations. Here $B^{z**}(x=\Delta x/2, y=\Delta y/2)$ and $B^{z**}(x=\Delta x/2, y=-\Delta y/2)$ are magnetic induction components that are obtained at the two endpoints of the right *x*-face. They are obtained by the application of a two-dimensional Riemann solver at the edges of the mesh; see Fig. 1. The factor $(1/12)$ in eqn. (3.5) is the analogue of a mass matrix. Because we have a Cartesian mesh with orthogonal bases, the mass matrix is diagonal. Also notice that the terms within angled brackets, i.e. terms with $\langle \ \rangle$, represent suitably high order line averages within a face; these terms with an angled bracket are to be obtained with a suitably high order quadrature along each face of the mesh. In this work, since the *z*-variation is suppressed, we use the well-known one-dimension Gauss-Legendre quadrature to carry out the facial integrals. One dimensional Riemann problems in the right face being considered will furnish the $B^{z*}(x=\Delta x/2, y)$ component of the magnetic induction field that is to be used in the angled brackets. These one-dimensional Riemann problems are solved at each of the quadrature points in the *x*-face.

In the upper *y*-face of the element shown in Fig. 1, i.e. in the face with $y=\Delta y/2$, we assert the second order accurate evolution of the *y*-component of the displacement vector to be of the form

$$D^y(x,t) = D_0^y(t) + D_x^y(t)\left(\frac{x}{\Delta x}\right) \tag{3.6}$$

The variation of the *y*-component of the electric displacement in eqn. (3.6) is also shown in Fig. 1. As before, we have suppressed the *z*-variation in the above equation because we are only interested in a two-dimensional scheme. Our trial functions are $\phi(x)=1$ and $\phi(x)=(x/\Delta x)$. Using $\hat{\mathbf{n}} = \hat{\mathbf{y}}$ and the test function $\phi(x)=1$, eqn. (3.2) then gives us

$$\frac{dD_0^y(t)}{dt} + \frac{1}{\mu}\frac{1}{\Delta x}\left[B^{z**}(x=\Delta x/2, y=\Delta y/2) - B^{z**}(x=-\Delta x/2, y=\Delta y/2)\right] = 0 \tag{3.7}$$

Using $\hat{\mathbf{n}} = \hat{\mathbf{y}}$ and the test function $\phi(x)=(x/\Delta x)$ in eqn. (3.2) we get

$$\frac{1}{12}\frac{dD_x^y(t)}{dt} + \frac{1}{\mu}\frac{1}{2\Delta x}\left[B^{z**}(x=\Delta x/2, y=\Delta y/2) + B^{z**}(x=-\Delta x/2, y=\Delta y/2)\right]$$
$$-\frac{1}{\mu}\frac{1}{\Delta x}\langle B^{z*}(x, y=\Delta y/2)\rangle = 0 \tag{3.8}$$

Here again $B^{z**}(x=\Delta x/2, y=\Delta y/2)$ and $B^{z**}(x=-\Delta x/2, y=\Delta y/2)$ are magnetic induction components that are obtained at the endpoints of the upper *y*-face. They are obtained by the application of a two-dimensional Riemann solver at the edges of the mesh; see Fig. 1. One dimensional Riemann problems in the upper face being considered will furnish the $B^{z*}(x, y=\Delta y/2)$ component of the magnetic induction field that is to be used in the angled brackets. These one-dimensional Riemann problems are solved at each of the quadrature points in the *y*-face. Eqns. (3.4) and (3.7) taken together also ensure that the mean electric displacement field components within the faces of the mesh preserve the constraint-preserving property at a discrete level. In other words, we retrieve the traditional Yee-type update.

Now let us turn our focus to eqn. (2.1). From Eqn. (2.5) we see that we will be interested in the evolution of the *z*-component of the magnetic induction. This component resides in the *xy*-faces in Fig. 1. Say we take $\hat{\mathbf{n}} = \hat{\mathbf{z}}$ in Fig. 1; in that case the area $A_n$ will be the far *xy*-face which contains the field component $D_z$ that we are interested in evolving via eqn. (2.5). We follow a procedure that is analogous to the one that gave us eqn. (3.2), however, this time we start with eqn. (2.1) to get

$$\frac{\partial}{\partial t}\left(\int_{A_n}(\hat{\mathbf{n}}\cdot\mathbf{B})\phi \, dA_n\right)+\frac{1}{\varepsilon}\int_{\partial A_n}(\phi\,\mathbf{D})\cdot d\vec{\ell}-\frac{1}{\varepsilon}\int_{A_n}\hat{\mathbf{n}}\cdot\left[(\nabla\phi)\times\mathbf{D}\right]dA_n=0 \qquad (3.9)$$

The above equation is true in general. However, let us examine what happens when we use $\hat{\mathbf{n}} = \hat{\mathbf{z}}$ in Fig. 1. In that case, the second term is an integral over the entire boundary of the $xy$-face of Fig. 1. This integral will pick up contributions from $D_x$ and $D_y$ at the boundary; and these contributions are obtained from a one-dimensional Riemann solver. Also realize that the third term in eqn. (3.9) requires that we have the components of $D_x$ and $D_y$ at all locations in the $xy$-face. In other words, even though eqns. (3.3) and (3.6) only give us the components of the electric displacement at the boundaries, we need a strategy for reconstructing the electric displacement vector field at all locations in the element. In other words, we are forced to the interesting realization that a volumetric reconstruction strategy for the electric displacement vector field that is consistent with the boundary values in eqns. (3.3) and (3.6) as well as the constraints in eqns. (2.3) and (2.4) is an essential ingredient in any DG scheme for CED. Such a second order, constraint-preserving reconstruction has been described in Section III of Balsara *et al*. [18] or Balsara *et al*. [19]. Eqn. (3.8) is our master equation that stems from Faraday's law. In the next few paragraphs we will show how it is to be used to design a DG scheme at second order.

Now let us be specific with respect to eqn. (3.9). In the near $z$-face of Fig. 1 we assert the second order accurate evolution of the $z$-component of the magnetic induction vector to be of the form

$$B^z(x,y,t) = B_0^z(t) + B_x^z(t)\left(\frac{x}{\Delta x}\right) + B_y^z(t)\left(\frac{y}{\Delta y}\right) \qquad (3.10)$$

The variation of the $z$-component of the magnetic induction in eqn. (3.10) is also shown in Fig. 1. Notice that suppressing the z-variation implies that both the near and far $z$-faces of Fig. 1 have the same $z$-component of the magnetic induction vector, thus ensuring that the constraint in eqn. (2.3) is always satisfied for the magnetic induction. (The far $z$-face in Fig. 1 is not shown.) Our trial functions are $\phi(x,y)=1$, $\phi(x,y)=(x/\Delta x)$ and $\phi(x,y)=(y/\Delta y)$. Using $\hat{\mathbf{n}}=\hat{\mathbf{z}}$ and the test function $\phi(x,y)=1$ in eqn. (3.9) then gives us

$$\frac{dB_0^z(t)}{dt}+\frac{1}{\varepsilon}\frac{1}{\Delta x}\Big[\langle D^{y*}(x=\Delta x/2,y)\rangle-\langle D^{y*}(x=-\Delta x/2,y)\rangle\Big]$$
$$-\frac{1}{\varepsilon}\frac{1}{\Delta y}\Big[\langle D^{x*}(x,y=\Delta y/2)\rangle-\langle D^{x*}(x,y=-\Delta y/2)\rangle\Big]=0 \qquad (3.11)$$

Using $\hat{\mathbf{n}}=\hat{\mathbf{z}}$ and the test function $\phi(x,y)=(x/\Delta x)$ in eqn. (3.9) also gives us

$$\frac{1}{12}\frac{dB_x^z(t)}{dt}+\frac{1}{\varepsilon}\frac{1}{2\Delta x}\Big[\langle D^{y*}(x=\Delta x/2,y)\rangle+\langle D^{y*}(x=-\Delta x/2,y)\rangle\Big]$$
$$-\frac{1}{\varepsilon}\frac{1}{\Delta y}\Big[\langle (x/\Delta x)D^{x*}(x,y=\Delta y/2)\rangle-\langle (x/\Delta x)D^{x*}(x,y=-\Delta y/2)\rangle\Big]-\frac{1}{\varepsilon}\frac{1}{\Delta x}\{D^{y}(x,y)\}=0 \qquad (3.12)$$

Using $\hat{\mathbf{n}}=\hat{\mathbf{z}}$ and the test function $\phi(x,y)=(y/\Delta y)$ in eqn. (3.9) further gives us

$$\frac{1}{12}\frac{dB_y^z(t)}{dt}+\frac{1}{\varepsilon}\frac{1}{\Delta x}\Big[\langle (y/\Delta y)D^{y*}(x=\Delta x/2,y)\rangle-\langle (y/\Delta y)D^{y*}(x=-\Delta x/2,y)\rangle\Big]$$
$$-\frac{1}{\varepsilon}\frac{1}{2\Delta y}\Big[\langle D^{x*}(x,y=\Delta y/2)\rangle+\langle D^{x*}(x,y=-\Delta y/2)\rangle\Big]+\frac{1}{\varepsilon}\frac{1}{\Delta y}\{D^{x}(x,y)\}=0 \qquad (3.13)$$

From the above three equations, note that angled brackets again represent suitably high order line averages in the edges that surround the *z*-face. Notice that the angled brackets in the above three equations only contain the electric displacements obtained from one-dimensional Riemann solvers. This is because the two-dimensional Riemann solver applied to the *x*-edges and *y*-edges of Fig. 1 reduces to a one-dimensional Riemann solver when the entire *z*-variation is suppressed. Also notice the introduction of curly brackets, i.e. { } , in eqns. (3.12) and (3.13). These curly brackets denote suitably high order area averages within the *z*-face. As always, they have to be obtained via a suitably high order two-dimensional quadrature formula. Alternatively, since $D^x(x,y)$ and $D^y(x,y)$ are expressed in terms of an orthogonal basis set, the curly brackets can usually be evaluated analytically on a Cartesian mesh. Eqns. (3.12) and (3.13) show very clearly that we should use the facial variation from eqns. (3.3) and (3.6) to obtain a second order, constraint-preserving reconstruction within the element for the electric displacements $D^x(x,y)$ and $D^y(x,y)$. This completes our description of the second order, P=1, DGTD scheme for CED.

Eqns. (3.11), (3.12) and (3.13) show something else that is also very interesting. Observe the last row of eqn. (2.5) and realize that it has the form of a traditional conservation law. Applying classical DG formulations to the last line of eqn. (2.5) would also yield eqns. (3.11), (3.12) and (3.13). This establishes a very nice consistency between our new DG-like schemes and classical DG schemes. In the limit where both are expected to yield the same result, they indeed do yield the same result!

**III.b) von Neumann Stability Analysis for Second Order, P=1, DGTD Scheme for Maxwell's Equations**

The von Neumann stability analysis of DG schemes can be done in one of two alternative ways. The first way follows the approach of Balsara and Käppeli [18] who convert the modal representation into a nodal representation. Liu *et al*. [35] also used a similar approach for analyzing DG schemes. In that approach, the DG equations are converted into a finite-difference-like formulation. The other approach, which we use here, retains the modal representation but endows it with a periodicity that is consistent with the Fourier modes.

Let us look at Fig. 1 and first focus on the electric displacement vector. In the right *x*-face we identify the modes $D_0^{x+}(t)$ and $D_y^{x+}(t)$ which pertain to the mean value of the *x*-component of the electric displacement and its slope in the *y*-direction. In the left *x*-face we can again identify the modes $D_0^{x-}(t)$ and $D_y^{x-}(t)$. In the upper *y*-face we identify the modes $D_0^{y+}(t)$ and $D_x^{y+}(t)$ which pertain to the mean value of the *y*-component of the electric displacement and its slope in the *x*-direction. In the lower *y*-face we can again identify the modes $D_0^{y-}(t)$ and $D_x^{y-}(t)$. We now assert a Fourier variation of the form $e^{i(k_x x + k_y y)}$ with wave numbers $k_x$ and $k_y$ on a uniform mesh with zones of size $\Delta x$ and $\Delta y$. Because we use periodic boundary conditions in our von Neumann stability analysis, the modes in the right and left *x*-faces are related; analogously, the modes in the upper and lower *y*-faces are also related. We can, therefore, write

$$D_0^{x-}(t) = D_0^{x+}(t) e^{-ik_x \Delta x} \quad ; \quad D_y^{x-}(t) = D_y^{x+}(t) e^{-ik_x \Delta x} \quad ;$$
$$D_0^{y-}(t) = D_0^{y+}(t) e^{-ik_y \Delta y} \quad ; \quad D_x^{y-}(t) = D_x^{y+}(t) e^{-ik_y \Delta y}$$
(3.14)

In light of eqn. (3.14), it would seem that each element has four genuinely independent pieces of data associated with the electric displacement, however, this is not true. The equations also satisfy a discrete divergence-free condition that can be written as

$$\frac{D_0^{x+}(t) - D_0^{x+}(t) e^{-ik_x \Delta x}}{\Delta x} + \frac{D_0^{y+}(t) - D_0^{y+}(t) e^{-ik_y \Delta y}}{\Delta y} = 0 \quad \Leftrightarrow \quad D_0^{y+}(t) = -D_0^{x+}(t) \frac{\Delta y}{\Delta x} \frac{1 - e^{-ik_x \Delta x}}{1 - e^{-ik_y \Delta y}} \quad (3.15)$$

We see, therefore, that associated with each element, we have only three truly independent modes for the electric displacement that participate in the von Neumann stability analysis. In light of eqn. (3.15), those modes are explicitly given by $D_0^{x+}(t)$, $D_y^{x+}(t)$ and $D_x^{y+}(t)$. Using eqns. (3.14) and (3.15), all electric displacement-related quantities in the von Neumann stability analysis can be written in terms of these three modes. This is easily illustrated in Fig. 2a; note however that the time-dependence has been suppressed in order to keep the equations on the figure manageable. Please note that eqn. (3.15) has not been introduced in Fig. 2a, with the result that it seems as if there are four independent modes, but we request the reader to imagine that $D_0^{y+}$ has always been replaced by $D_0^{x+}$ with the use of eqn. (3.15).

Let us look at Fig. 1 and first focus on the magnetic induction vector in the *z*-faces of Fig. 1. The near and far *z*-faces in Fig. 1 have identical variation. As a result, the z-face of each element, like the element shown in Fig. 1, will have three independent modes. For Fig. 1, we identify these modes as $B_0^{z+}(t)$, $B_x^{z+}(t)$ and $B_y^{z+}(t)$. The magnetic induction in all neighboring faces can be related to these three modes via Fourier variation analogous to eqn. (3.14) when carrying out a von Neumann stability analysis. This is easily illustrated in Fig. 2b; note however that the time-dependence has been suppressed in order to keep the equations on the figure manageable.

Using the Fourier dependence from eqns. (3.14) and (3.15), the constraint-preserving reconstruction for the electric displacement vector field in each of the nine zones in Fig. 2a can be expressed in terms of our three truly independent modes for the electric displacement. Likewise, using the Fourier dependence in Fig. 2b, we can express the magnetic induction in each of the nine zones of Fig. 2b in terms of our three truly independent modes for the magnetic induction. The equations that we are interested in are given by eqns. (3.4), (3.5), (3.8), (3.11), (3.12) and (3.13). (Recall that because of eqn. (3.15), we can bypass eqn. (3.7).) Given the linearity of the DG scheme for CED, we can write the time rate of change of our six truly independent modes as the following linear system of ODEs:-

$$\frac{d}{dt}\begin{pmatrix} D_0^{x+}(t) \\ D_y^{x+}(t) \\ D_x^{y+}(t) \\ B_0^{z+}(t) \\ B_x^{z+}(t) \\ B_y^{z+}(t) \end{pmatrix} = \begin{pmatrix} A_{11} & A_{12} & A_{13} & A_{14} & A_{15} & A_{16} \\ A_{21} & A_{22} & A_{23} & A_{24} & A_{25} & A_{26} \\ A_{31} & A_{32} & A_{33} & A_{34} & A_{35} & A_{36} \\ A_{41} & A_{42} & A_{43} & A_{44} & A_{45} & A_{46} \\ A_{51} & A_{52} & A_{53} & A_{54} & A_{55} & A_{56} \\ A_{61} & A_{62} & A_{63} & A_{64} & A_{65} & A_{66} \end{pmatrix} \begin{pmatrix} D_0^{x+}(t) \\ D_y^{x+}(t) \\ D_x^{y+}(t) \\ B_0^{z+}(t) \\ B_x^{z+}(t) \\ B_y^{z+}(t) \end{pmatrix} \quad (3.16)$$

The 36 matrix elements in eqn. (3.16) depend only on $(k_x \Delta x)$, $(k_y \Delta y)$ and the speed of light "$c$". They are explicitly given in Appendix A. We refer to the $6 \times 6$ matrix in the above equation as $\mathbf{A}$. At any time "$t$" we also define the vector $\mathbf{V}(t) = \left(D_0^{x+}(t), D_y^{x+}(t), D_x^{y+}(t), B_0^{z+}(t), B_x^{z+}(t), B_y^{z+}(t)\right)^T$.

We then discretize eqn. (3.16) in time with an explicit $m$-stage Runge-Kutta scheme having a timestep $\Delta t$ of the form

$$\mathbf{V}^{(0)} = \mathbf{V}(t^n)$$
$$\mathbf{V}^{(i)} = \sum_{k=0}^{i-1} \left( \alpha_{i,k} \mathbf{I} + \Delta t \beta_{i,k} \mathbf{A} \right) \mathbf{V}^{(k)} \quad \text{for } i = 1,...,m \quad (3.17)$$
$$\mathbf{V}(t^{n+1}) = \mathbf{V}^{(m)}$$

Here "$\mathbf{I}$" is the identity matrix. The expressions for the coefficients $\alpha_{i,k}$ and $\beta_{i,k}$ can be found in Gottlieb *et al*. [31] and also Spiteri and Ruuth [38], [39]. Given the linearity of our DG scheme, we can write the time update as

$$\mathbf{V}(t^{n+1}) = \mathbf{G}\, \mathbf{V}(t^n) \quad (3.18)$$

Here "**G**" is known as the amplification matrix of the scheme. It depends on the coefficients of the Runge-Kutta scheme, on the timestep $\Delta t$ and the matrix "**A**" from eqn. (3.16). For the second order SSP-RK scheme we can write the amplification matrix as

$$\mathbf{G} = \mathbf{I} + \Delta t \mathbf{A} + \frac{\Delta t^2}{2} \mathbf{A}^2 \tag{3.19}$$

Likewise, for the third order SSP-RK scheme we can write the amplification matrix as

$$\mathbf{G} = \mathbf{I} + \Delta t \mathbf{A} + \frac{\Delta t^2}{2} \mathbf{A}^2 + \frac{\Delta t^3}{3} \mathbf{A}^3 \tag{3.20}$$

In Section V we will use this amplification matrix to devise our von Neumann stability analysis. This completes our description of the mathematics associated with the von Neumann stability analysis at second order.

**IV) Third Order, P=2, DGTD Scheme for Maxwell's Equations**

The previous section has shown us in great detail how to obtain update equations for the DG-like scheme that relies on the curl-type master equations, i.e. eqns. (3.2) and (3.9). That same style of obtaining update equations can be extended to the third order, i.e. P=2, DGTD scheme for CED.

The analogue of eqn. (3.3) at third order can be written as

$$D^x(y,t) = D_0^x(t) + D_y^x(t)\left(\frac{y}{\Delta y}\right) + D_{yy}^x(t)\left(\left(\frac{y}{\Delta y}\right)^2 - \frac{1}{12}\right) \tag{4.1}$$

Owing to the orthogonal nature of the basis functions, the update equations for $D_0^x(t)$ and $D_y^x(t)$ are still given by eqns. (3.4) and (3.5). The only caveat is that for third order accuracy, the quadrature formulae in those two equations should also be of higher order. The update equation for $D_{yy}^x(t)$ is obtained by using $\hat{\mathbf{n}} = \hat{\mathbf{x}}$ and the test function $\phi(y) = \left((y/\Delta y)^2 - 1/12\right)$ in eqn. (3.2) to get

$$\frac{1}{180}\frac{dD_{yy}^{x}(t)}{dt} - \frac{1}{\mu}\frac{1}{6\Delta y}\left[B^{z**}\left(x=\Delta x/2, y=\Delta y/2\right) - B^{z**}\left(x=\Delta x/2, y=-\Delta y/2\right)\right]$$
$$+ \frac{1}{\mu}\frac{2}{\Delta y}\left\langle (y/\Delta y)B^{z*}\left(x=\Delta x/2, y\right)\right\rangle = 0 \qquad (4.2)$$

The analogue of eqn. (3.6) at third order can be written as

$$D^{y}(x,t) = D_{0}^{y}(t) + D_{x}^{y}(t)\left(\frac{x}{\Delta x}\right) + D_{xx}^{y}(t)\left(\left(\frac{x}{\Delta x}\right)^{2} - \frac{1}{12}\right) \qquad (4.3)$$

As before, the update equations for $D_{0}^{y}(t)$ and $D_{x}^{y}(t)$ are still given by eqns. (3.7) and (3.8). The update equation for $D_{xx}^{y}(t)$ is obtained by using $\hat{\mathbf{n}} = \hat{\mathbf{y}}$ and the test function $\phi(x) = \left((x/\Delta x)^{2} - 1/12\right)$ in eqn. (3.2) to get

$$\frac{1}{180}\frac{dD_{xx}^{y}(t)}{dt} + \frac{1}{\mu}\frac{1}{6\Delta x}\left[B^{z**}\left(x=\Delta x/2, y=\Delta y/2\right) - B^{z**}\left(x=-\Delta x/2, y=\Delta y/2\right)\right]$$
$$- \frac{1}{\mu}\frac{2}{\Delta x}\left\langle (x/\Delta x)B^{z*}\left(x, y=\Delta y/2\right)\right\rangle = 0 \qquad (4.4)$$

The analogue of eqn. (3.10) at third order can be written as

$$B^{z}(x,y,t) = B_{0}^{z}(t) + B_{x}^{z}(t)\left(\frac{x}{\Delta x}\right) + B_{y}^{z}(t)\left(\frac{y}{\Delta y}\right)$$
$$+ B_{xx}^{z}(t)\left(\left(\frac{x}{\Delta x}\right)^{2} - \frac{1}{12}\right) + B_{yy}^{z}(t)\left(\left(\frac{y}{\Delta y}\right)^{2} - \frac{1}{12}\right) + B_{xy}^{z}(t)\left(\frac{x}{\Delta x}\right)\left(\frac{y}{\Delta y}\right) \qquad (4.5)$$

As before, the update equations for $B_{0}^{z}(t)$, $B_{x}^{z}(t)$ and $B_{y}^{z}(t)$ are still given by eqns. (3.11), (3.12) and (3.13). The update equation for $B_{xx}^{z}(t)$ is obtained by using $\hat{\mathbf{n}} = \hat{\mathbf{z}}$ and the test function $\phi(x,y) = \left((x/\Delta x)^{2} - 1/12\right)$ in eqn. (3.9) to get

$$\frac{1}{180}\frac{dB_{xx}^z(t)}{dt}+\frac{1}{\varepsilon}\frac{1}{6\Delta x}\Big[\langle D^{y*}(x=\Delta x/2,y)\rangle-\langle D^{y*}(x=-\Delta x/2,y)\rangle\Big]$$
$$-\frac{1}{\varepsilon}\frac{1}{\Delta y}\Big[\langle\big((x/\Delta x)^2-1/12\big)D^{x*}(x,y=\Delta y/2)\rangle-\langle\big((x/\Delta x)^2-1/12\big)D^{x*}(x,y=-\Delta y/2)\rangle\Big] \quad (4.6)$$
$$-\frac{1}{\varepsilon}\frac{2}{\Delta x}\{(x/\Delta x)D^y(x,y)\}=0$$

The update equation for $B_{yy}^z(t)$ is obtained by using $\hat{\mathbf{n}}=\hat{\mathbf{z}}$ and the test function $\phi(x,y)=\big((y/\Delta y)^2-1/12\big)$ in eqn. (3.9) to get

$$\frac{1}{180}\frac{dB_{yy}^z(t)}{dt}+\frac{1}{\varepsilon}\frac{1}{\Delta x}\Big[\langle\big((y/\Delta y)^2-1/12\big)D^{y*}(x=\Delta x/2,y)\rangle-\langle\big((y/\Delta y)^2-1/12\big)D^{y*}(x=-\Delta x/2,y)\rangle\Big]$$
$$-\frac{1}{\varepsilon}\frac{1}{6\Delta y}\Big[\langle D^{x*}(x,y=\Delta y/2)\rangle-\langle D^{x*}(x,y=-\Delta y/2)\rangle\Big]+\frac{1}{\varepsilon}\frac{2}{\Delta y}\{(y/\Delta y)D^x(x,y)\}=0$$

$$(4.7)$$

The update equation for $B_{xy}^z(t)$ is obtained by using $\hat{\mathbf{n}}=\hat{\mathbf{z}}$ and the test function $\phi(x,y)=(x/\Delta x)(y/\Delta y)$ in eqn. (3.9) to get

$$\frac{1}{144}\frac{dB_{xy}^z(t)}{dt}+\frac{1}{\varepsilon}\frac{1}{2\Delta x}\Big[\langle(y/\Delta y)D^{y*}(x=\Delta x/2,y)\rangle+\langle(y/\Delta y)D^{y*}(x=-\Delta x/2,y)\rangle\Big]$$
$$-\frac{1}{\varepsilon}\frac{1}{2\Delta y}\Big[\langle(x/\Delta x)D^{x*}(x,y=\Delta y/2)\rangle+\langle(x/\Delta x)D^{x*}(x,y=-\Delta y/2)\rangle\Big] \quad (4.8)$$
$$-\frac{1}{\varepsilon}\frac{1}{\Delta x}\{(y/\Delta y)D^y(x,y)\}+\frac{1}{\varepsilon}\frac{1}{\Delta y}\{(x/\Delta x)D^x(x,y)\}=0$$

The curly brackets in the above three equations require a suitably high order areal averaging of appropriate moments of the reconstructed electric displacement vector field within the element. Such a third order, constraint-preserving reconstruction has been described in Section III of Balsara *et al*. [20]. (The Appendix of that same paper even describes an analogous fourth order reconstruction, which could be useful in a fourth order accurate DGTD scheme.) This completes our description of the third order, P=2, DGTD scheme for CED.

The von Neumann stability analysis at third order proceeds analogously to the one in Section III.b for the second order case. We assert a Fourier variation between the facial variables, just as in eqn. (3.14). Just as in eqn. (3.15), $D_0^{y+}(t)$ can be expressed in terms of $D_0^{x+}(t)$. As a result, the five truly independent modes for the electric displacement that participate in the von Neumann stability analysis are explicitly given by $D_0^{x+}(t)$, $D_y^{x+}(t)$, $D_{yy}^{x+}(t)$, $D_x^{y+}(t)$ and $D_{xx}^{y+}(t)$. Likewise, the six truly independent modes for the magnetic induction that participate in the von Neumann stability analysis are explicitly given by $B_0^{z+}(t)$, $B_x^{z+}(t)$, $B_y^{z+}(t)$, $B_{xx}^{z+}(t)$, $B_{yy}^{z+}(t)$ and $B_{xy}^{z+}(t)$. Given the linearity of the DG scheme for CED, we can write the time rate of change of our eleven truly independent modes as a linear system of ODEs. Analogously to eqn. (3.16), the ODE system is now characterized by a $11 \times 11$ matrix. It is not worthwhile to write this matrix out explicitly, especially since it is obtained as an output from a computer algebra system. While a second order in time SSP-RK scheme cannot be used with a third order in space DG scheme, the amplification matrix for a third order in time SSP-RK scheme with the third order spatial DG discretization from this section is given by eqn. (3.20). This amplification matrix "**G**" will also be an $11 \times 11$ matrix. This completes our description of the mathematics associated with the von Neumann stability analysis at third order.

## V) Results from the von Neumann Stability Analysis of DGTD Schemes for CED

Because we have carried out our von Neumann stability analysis for the full DGTD schemes for CED, we are now in a position to extract a wealth of insights from the analysis. The most important information concerns the stability limit of the Runge-Kutta timestepping strategy. In other words, we wish to find the largest possible CFL number for a DGTD scheme that uses a certain order of spatial discretization along with a certain order of temporal discretization. This information is provided in Sub-section V.a for DGTD schemes that are second, third and fourth order accurate. Sub-section V.b also provides analogous information for PNPM schemes for CED. The motivation for this class of PNPM schemes, which are close relatives of the DGTD schemes, was discussed in the Introduction and will also be amplified in that Sub-section. The von Neumann stability analysis can also give us important insights into the wave-propagation characteristics of our DGTD and PNPM schemes for CED. Ideally, we would like waves to be represented over a small enough number of zones and we would, nevertheless, like those waves to propagate with minimal dissipation and dispersion. Moreover, we would like this wave propagation to be as isotropic as possible on the computational mesh. In Sub-section V.c we document how well waves propagate at various angles to the mesh for our second, third and fourth order DGTD schemes for CED. Sub-section V.d documents how well waves propagate at various angles to the mesh for our second third and fourth order PNPM schemes for CED. This work also enables us to quantify the number of zones that should be encompassed by an electromagnetic wave if we want it to propagate on a computational mesh with dispersion and dissipation that are held below user-specified tolerances.

## V.a) Stable CFL Numbers for Constraint-Preserving DGTD Schemes for CED

Eqns. (3.16), (3.19) and (3.20) clearly show that even a second order DGTD scheme can indeed produce a rather large amplification matrix. The sizes of these amplification matrices only increase with increasing order of the DGTD scheme. To recap from Sections III, IV and Appendix B, a second order DGTD scheme will yield a $6\times 6$ amplification matrix; a third order DGTD scheme will produce an $11\times 11$ amplification matrix and a fourth order DGTD scheme results in a $19\times 19$ amplification matrix. The eigenstructure of such large matrices cannot be analyzed analytically. As a result, our deduction of the CFL number is numerically motivated. We realize that the pair of normalized wavenumbers $(k_x \Delta x, k_y \Delta y)$ lie in the two-dimensional domain given by $[-\pi, \pi] \times [-\pi, \pi]$. We subdivide this domain into a grid of $201\times 201$ cells. Each cell, therefore, identifies a particular wavenumber pair given by $(k_x \Delta x, k_y \Delta y)$, or equivalently, a particular direction in which electromagnetic waves propagate on the mesh. An automated computer program examines each of the $201\times 201$ cells. For each cell, the amplification matrix is evaluated for the entire DGTD scheme for increasing values of $(c\Delta t/\Delta x)$ (or equivalently $(c\Delta t/\Delta y)$). The largest value of $(c\Delta t/\Delta x)$ for which we obtain an amplification matrix with eigenvalues that are bounded by $[-1, 1]$ is deemed to be the stable CFL number for that particular pair of wavenumbers $\left(k_x \Delta x, k_y \Delta y\right)$. The final, stable CFL number is the smallest of the CFL numbers obtained for wave propagation in all possible directions.

The narrative from the previous paragraphs enables us to make a table that is suitable for practical use. Table I shows the maximal CFL number for DGTD schemes with various orders of spatial accuracy that are used in conjunction with Runge-Kutta schemes with various orders of temporal accuracy. A dash in Table I indicates that the scheme is unstable. Observe that the limiting CFL in Table I is very competitive (albeit slightly different) from the limiting CFLs for DG schemes from Cockburn and Shu [27]. Please see Table 2.2 from Cockburn and Shu [27]. Recall that the limiting CFL analysis of Cockburn and Shu [27] is strictly one-dimensional whereas our results are multidimensional. This suggests that a full-fledged DGTD scheme for constraint-preserving CED will have CFL numbers that are competitive with conventional RKDG schemes. This bodes well for the construction of constraint-preserving DGTD schemes for CED.

**Table I shows the limiting CFL number for a large number of possible DGTD schemes for CED. The table shows spatial order of accuracy of DG schemes in the horizontal direction and temporal order of accuracy of the Runge-Kutta timestepping in the vertical direction.**

|            | P=0    | P=1    | P=2    | P=3    |
|------------|--------|--------|--------|--------|
| RK1        | 0.5000 | ______ | ______ | ______ |
| SSP-RK2    | 0.5000 | 0.2500 | ______ | ______ |
| SSP-RK3    | 0.6282 | 0.3141 | 0.1623 | 0.1095 |
| SSP-RK(5,4)| 1.3329 | 0.5895 | 0.3373 | 0.2153 |

**V.b) Stable CFL Numbers for Constraint-Preserving PNPM Schemes for CED**

Scanning Table I horizontally, we see that the CFL number decreases with increasing order of the scheme. We would like to obtain larger CFL numbers even for higher order schemes. PNPM schemes, which are close cousins of DG schemes, are one way of achieving that goal. In Dumbser *et al*. [29] we showed that PNPM schemes occupy a conceptual space that is intermediate between WENO and DGTD schemes. For example, the P0PM scheme can be a TVD/WENO scheme that evolves just the mean value and reconstructs all the higher moments. Furthermore, a P1PM scheme updates the mean value and its first moment while reconstructing the higher order moments. Indeed, the first moments are updated using the same update equations that are shown in Section III, with the only exception that the order of accuracy of the quadrature should keep up with the $M^{th}$ order of accuracy of the scheme. Likewise, and again by way of example, a P2PM scheme updates the mean value as well as the first and second moments, while reconstructing all the higher order moments. The update equations for the second moments are shown in Section IV, with the same caveat about the quadrature. The $M^{th}$ order accurate, i.e. P=M, DGTD schemes can, therefore, be thought of as PMPM schemes and the WENO schemes can be thought of as P0PM schemes. Please see Fig. 1 of Dumbser *et al*. [29] for further insights on PNPM schemes. It was noted in Dumbser *et al*. [29] that P1PM schemes offer accuracies that are almost as good as the accuracies of the PMPM schemes. However, they permit a substantially larger CFL. A look at Table I shows that the CFL of DGTD schemes decreases with increasing order of accuracy. This limits the utility of very high order DGTD schemes. The PNPM schemes were shown in Dumbser *et al*. [29] to overcome this issue of progressively smaller CFL with increasing order of the DG scheme. For the case of CED, Table II shows the limiting CFL number for several possible PNPM schemes for CED. The P0P2 scheme is just a third order WENO scheme; the P1P2 and P1P3 schemes are Hermite WENO schemes at third and fourth orders. Table II clearly shows us that the P0P1 and P0P2 schemes have CFL numbers that are comparable to the corresponding P=0 DG schemes from Table I. The P1P2 and P1P3 schemes have CFL numbers that are comparable to the P=1 DG schemes. This mirrors the trend found in Dumbser *et al*. [29] for finite volume-type methods and extends the result to constraint-preserving CED.

**Table II shows the limiting CFL number for a large number of possible PNPM schemes for CED. The table shows spatial order of accuracy of PNPM schemes in the horizontal direction and temporal order of accuracy of the Runge-Kutta timestepping in the vertical direction.**

|            | P0P1   | P0P2   | P1P2   | P0P3   | P1P3   |
|------------|--------|--------|--------|--------|--------|
| SSP-RK2    | 0.5000 | ____   | ____   | ____   | ____   |
| SSP-RK3    | 0.6282 | 0.9423 | 0.3141 | 0.9423 | 0.3371 |
| SSP-RK(5,4)| 1.3329 | 1.6776 | 0.5895 | 1.5378 | 0.5801 |

**V.c) Analyzing the Propagation of Electromagnetic Waves for DGTD Schemes**

In CED we would like electromagnetic waves to propagate as isotropically as possible relative to the computational mesh. They should propagate with speeds that are as close to the speed of light as possible. Since electromagnetic waves are not dissipated as they propagate in perfect insulators with uniform dielectric properties, we want the dissipation of the numerical scheme to be as small as possible. Moreover, we would like all these desirable properties to hold for electromagnetic waves with a wavelength that spans as few zones as possible. We are helped in this regard by the fact that our DGTD scheme is linear, at least when limiters are not used. The von Neumann stability analysis can give us an abundance of insights with regard to electromagnetic wave propagation on a Cartesian computational mesh. Fig. 4.2 from Taflove and Hagness [44] presents a von Neumann analysis-based study of electromagnetic wave propagation in FDTD, thereby providing the motivation for an analogous study in this section.

Operationally, we work on a Cartesian mesh with $\Delta x = \Delta y$. We choose different values for $\mathbf{k} = k_x \hat{\mathbf{x}} + k_y \hat{\mathbf{y}}$ while keeping the product $|\mathbf{k}|\Delta x$ fixed. We also choose a value for the CFL number. In the results shown here the CFL number was held to be 95% of the maximal allowable CFL number for the scheme shown. For each direction of electromagnetic wave propagation, the von Neumann stability analysis then gives us an amplification factor which is the largest normed value of the eigenvector of the amplification matrix. It also gives us a phase speed for the propagation of the waves. Ideally, we want the amplification factor to be as close to unity as possible. We also want the phase speed of the waves to be as close to the speed of light as possible.

Fig. 3a shows the amplification factor for wave propagation in various directions relative to the mesh for waves that have a wavelength of five zones. The green, red, cyan and blue curves show the results for the second order P=1 DGTD scheme, the third order P=2 DGTD scheme, the fourth order P=3 DGTD scheme and the Yee scheme respectively. The temporal accuracy of the Runge-Kutta timestepping matches the spatial accuracy of the DG scheme. Fig. 3b shows the phase velocity, normalized to unity, for the same four schemes using the same color coding. The results correspond to a CFL that is 95% of the maximum. Let us focus first on Fig. 3a which pertains to dissipation. We see that the second order, P=1, DGTD scheme has an amplification factor that is smaller than unity by a substantial amount. As a result, it will be rather dissipative when waves with a wavelength of five zones are represented on the computational mesh. The third order, P=2, DGTD scheme already shows a considerably reduced dissipation. The fourth order, P=3, DGTD scheme is almost free of dissipation and comparable to the Yee scheme which is fully non-dissipative on account of it being a symplectic scheme. Now let us focus on Fig. 3b which pertains to dispersion. Both the second order, P=1, DGTD scheme as well as the second order Yee scheme show themselves to be rather dispersive. By contrast, the third order, P=2, DGTD scheme shows substantially improved dispersion which lies much closer to the ideal value of unity for all directions of wave propagation. The fourth order, P=3, DGTD scheme shows an even more appealing result because the dispersion is practically perfect. We see that with increasing order of accuracy, the DGTD schemes become closer to the ideal limit in their dissipation as well as their dispersion. Table III quantifies the dissipation and the dispersion when the wavelength of the electromagnetic radiation spans five zones. To quantify the dissipation, we identify the range of values assumed by the amplification factor when waves propagate in any direction relative to the mesh in Fig. 3a. This range is denoted by $[\lambda_{min}, \lambda_{max}]$. To quantify the dispersion, we quantify the maximum swing away from unity in the normalized phase velocity when waves propagate in any direction relative to the mesh in Fig. 3b. The swing in the phase away from unity is denoted by $|\text{Phase} - 1|$.

**Table III quantifies the dissipation and the dispersion of various DGTD schemes when the wavelength of the electromagnetic radiation spans five zones.**

| Scheme | $\lambda_{min}$ | $\lambda_{max}$ | $|\text{Phase} - 1|$ |
|---|---|---|---|
| P=1 DGTD | 0.98900738 | 0.99348345 | 3.9209e-02 |
| P=2 DGTD | 0.99475966 | 0.99986534 | 8.1188e-03 |
| P=3 DGTD | 0.99972425 | 0.99999870 | 1.3655e-03 |
| 2nd Order Yee | 1.0 | 1.0 | 3.8321e-02 |

Fig. 4a shows the amplification factor for wave propagation in various directions relative to the mesh for waves that have a wavelength of ten zones. The green, red, cyan and blue curves show the results for the second order P=1 DGTD scheme, the third order P=2 DGTD scheme, the fourth order P=3 DGTD scheme and the Yee scheme respectively. The temporal accuracy of the Runge-Kutta timestepping matches the spatial accuracy of the DG scheme. Fig. 4b shows the phase velocity, normalized to unity, for the same four schemes using the same color coding. The results correspond to a CFL that is 95% of the maximum. Focusing on Fig. 4a, we see that the second order, P=1, DGTD scheme still shows some very small amount of relic dissipation. However, the third order, P=2, DGTD scheme is very close to perfect. The fourth order, P=3, DGTD scheme is virtually indistinguishable from the Yee scheme in its dissipation properties. Focusing on Fig. 4b, we see that the second order, P=1, DGTD scheme is somewhat better than the Yee scheme in its dispersive properties. This is because it evolves the mean value as well as its linear variation – as opposed to the Yee scheme which only evolves the mean value. Fig. 4b also shows the very pleasing result that the third and fourth order DGTD schemes are both practically perfect in their lack of dispersion. This is a very desirable property, especially when we want electromagnetic waves to propagate over thousands of zones in a large simulation. (For certain timing applications, like the design of a GPS system, the phase error can indeed be of paramount importance.) We see that when we allow ten zones per wavelength, the third and fourth order DGTD schemes have come extremely close to the ideal limit both in their dissipative and their dispersive properties. Table IV, which is analogous to Table III, quantifies the dissipation and the dispersion when the wavelength of the electromagnetic radiation spans ten zones as shown in Fig. 4a and Fig. 4b.

**Table IV quantifies the dissipation and the dispersion of various DGTD schemes when the wavelength of the electromagnetic radiation spans ten zones.**

| Scheme | $\lambda_{min}$ | $\lambda_{max}$ | $|\text{Phase} - 1|$ |
|---|---|---|---|
| P=1 DGTD | 0.99913670 | 0.99956291 | 5.8649e-03 |
| P=2 DGTD | 0.99973145 | 0.99999505 | 5.4317e-04 |
| P=3 DGTD | 0.99999469 | 0.99999999 | 1.2099e-04 |
| 2nd Order Yee | 1.0 | 1.0 | 9.1637e-03 |

**V.d) Analyzing the Propagation of Electromagnetic Waves for PNPM Schemes**

Fig. 5 is analogous to Fig. 3, except that it pertains to the wave propagation at various angles for order P0P1, P0P2, P1P2 and P2P2 schemes for CED. The waves span five zones. The P2P2 scheme is just the P=2 DGTD scheme and is shown for reference. The vertical scales in Fig. 5 are different from Fig. 3. The P0P1 and P0P2 schemes are second and third order WENO schemes. We see that when the wave spans just five zones, there are deficiencies in these schemes because they try to reconstruct all the moments at each timestep. Comparison of the third order P1P2 and P2P2 schemes shows a very interesting result. We see that the P1P2 scheme provides almost the same high quality of wave propagation as the P2P2 scheme. This is because most of the variation within a zone is carried by the linear modes, which are indeed evolved in the P1P2 scheme. Let us consider the maximal CFL for both schemes when they are used along with a third order in time SSP-RK3 timestepping strategy. This data is available from Tables I and II. We see that the P1P2 scheme sustains a maximal CFL of 0.3141, which is considerably larger than the maximal CFL of 0.1623 for the P2P2 scheme.

Fig. 6 is also analogous to Figs. 5 and 3 because it shows the wave propagation at various angles for fourth order P0P3, P1P3 and P3P3 schemes for CED. The waves span five zones. The P3P3 scheme is just the P=3 DGTD scheme and is shown for reference. The vertical scales in Fig. 6 are different from Figs. 5 and 3. These fourth order PNPM schemes repeat the trends that we saw for the third order PNPM schemes in Fig. 5. We see that the fourth order P0P3 schemes have very good wave propagation properties on account of the fact that they are fourth order, however, they are not competitive with the P1P3 and P3P3 schemes. We also see the following interesting trend from the previous figure repeated in this figure:- We see that the P1P3 scheme provides competitively high quality of wave propagation as the P3P3 scheme. This is because most of the variation within a zone is carried by the linear modes, which are indeed evolved in the P1P3 scheme. Let us consider the maximal CFL for both schemes when they are used along with a fourth order in time SSP-RK(5,4) timestepping strategy. This data is available from Tables I and II. We see that the P1P3 scheme sustains a maximal CFL of 0.5895, which is considerably larger than the maximal CFL of 0.2153 for the P3P3 scheme.

Table V is culled from Figs. 5 and 6 and quantifies the dissipation and dispersion characteristics of various second, third and fourth order PNPM schemes when the wavelength of the electromagnetic radiation spans five zones. We urge the reader to compare results for P1P2 schemes from Table V with the similar results for P2P2 from Table III to appreciate that the two schemes are closely competitive. Likewise, we urge the reader to compare results from P1P3 schemes from Table V with analogous results for P3P3 from Table III to again appreciate that the two schemes are very competitive.

**Table V quantifies the dissipation and the dispersion of various PNPM schemes when the wavelength of the electromagnetic radiation spans five zones.**

| Scheme | $\lambda_{min}$ | $\lambda_{max}$ | $|\text{Phase} - 1|$ |
|---|---|---|---|
| P0P1 | 0.89000737 | 0.94364423 | 9.8062e-02 |
| P0P2 | 0.81973596 | 0.87364593 | 5.2768e-02 |
| P1P2 | 0.99092865 | 0.99802452 | 4.3205e-03 |
| P0P3 | 0.85975280 | 0.90222802 | 3.7129e-02 |
| P1P3 | 0.99458508 | 0.99707289 | 3.7463e-03 |

Figs. 5 and 6 have given us several useful insights into the operation of constraint-preserving PNPM schemes when the wavelength of the electromagnetic radiation spans a mere five zones. In most applications, the waves would span more zones. We wish to see if the trends that we found in Figs. 5 and 6 and encapsulated in Table V will carry over when the wavelength of the electromagnetic radiation spans a larger number of zones. Let us first focus on second and third order PNPM schemes. Fig. 7 is analogous to Fig. 4, except that it pertains to the wave propagation in various angles for P0P1, P0P2, P1P2 and P2P2 schemes for CED. In this figure the waves span ten zones. We see that all the trends that we found in Fig. 5 are indeed repeated here. The primary difference is that the dispersion and dissipation of the second and third order PNPM schemes experiences a dramatic improvement in wave propagation when the electromagnetic radiation spans ten zones instead of five. Now let us shift attention to fourth order PNPM schemes. Fig. 8 is also analogous to Figs. 7 and 4 because it shows the wave propagation at various angles for fourth order P0P3, P1P3 and P3P3 schemes for CED. In this figure the waves span ten zones. Again we see that all the trends that we found in Fig. 6 are indeed repeated here. Notice though that fourth order schemes are intrinsically quite accurate. As a result, we don't see quite as dramatic improvement in wave propagation for fourth order schemes when the electromagnetic radiation spans ten zones instead of five.

Table VI is culled from Figs. 7 and 8 and quantifies the dissipation and dispersion characteristics of various second, third and fourth order PNPM schemes when the wavelength of the electromagnetic radiation spans ten zones. We urge the reader to compare results for P1P2 schemes from Table VI with the similar results for P2P2 from Table IV to appreciate that the two schemes are closely competitive. Likewise, we urge the reader to compare results from P1P3 schemes from Table VI with analogous results for P3P3 from Table IV to again appreciate that the two schemes are very competitive.

**Table VI quantifies the dissipation and the dispersion of various PNPM schemes when the wavelength of the electromagnetic radiation spans ten zones.**

| Scheme | $\lambda_{min}$ | $\lambda_{max}$ | $|\text{Phase} - 1|$ |
| --- | --- | --- | --- |
| P0P1 | 0.99207109 | 0.99640157 | 4.0716e-02 |
| P0P2 | 0.98538641 | 0.99061259 | 2.3023e-03 |
| P1P2 | 0.99945071 | 0.99992869 | 2.7245e-04 |
| P0P3 | 0.99690890 | 0.99814612 | 2.0006e-03 |
| P1P3 | 0.99988990 | 0.99994704 | 1.4516e-04 |

## VI) Numerical Tests

Since we are primarily interested in dissipation and dispersion-free propagation of electromagnetic waves, we have focused on wave propagation in this section. All the numerical tests were done with DGTD and PNPM schemes where the temporal accuracy matched the spatial accuracy. This test is drawn from Balsara *et al*. [19], where it has been described in great detail. test problem consists of a plane polarized electromagnetic wave propagating in a vacuum along the north-east diagonal of a two-dimensional Cartesian mesh spanning $[-0.5, 0.5] \times [-0.5, 0.5]$ meter. Periodic boundary conditions were enforced. The magnetic induction was initialized using a magnetic vector potential given by

$$\mathbf{A}(x, y, z, t) = \frac{1}{2\pi} \sin\left[2\pi\left(x + y - \sqrt{2}ct\right)\right] \hat{y}$$

and the components of the magnetic induction vector were obtained at the zone faces by using the relationship $\mathbf{B} = \nabla \times \mathbf{A}$. The electric displacement was initialized using an electric vector potential given by

$$\mathbf{C}(x,y,z,t) = -\frac{1}{2\pi\sqrt{2}}\sin\left[2\pi\left(x+y-\sqrt{2}ct\right)\right]\hat{z}$$

and the components of the electric displacement vector were obtained at the zone faces by using the relationship $\mathbf{D} = c\varepsilon_0(\nabla \times \mathbf{C})$ where $c$ is the speed of light in free space and $\varepsilon_0 = 8.85 \times 10^{-12}$ F/m is the free space permittivity. With these analytical forms in hand, it is possible to evaluate the accuracy of the solution at any later time if it is set up correctly at the initial time on the mesh. For this EM field, we choose the wavelength to be 1 meter. The problem was run to a final time of 2.3587 nano-second on the computational mesh.

Tables VII, VIII and IX show the accuracy analysis for the second, third and fourth order DGTD schemes. The errors and accuracy in the y-component of the electric displacement vector and z-component of the magnetic induction are shown at the last time point in the simulation. We see that the schemes meet their design accuracies. It is also noteworthy that on the $8^2$ zone mesh the fourth order scheme is almost thirty times more accurate than the second order scheme! We also see that the schemes meet their design accuracies on resolution-starved meshes; in other words, going from the $8^2$ zone mesh to the $16^2$ zone mesh we see that the design accuracy is indeed realized. This sub-cell resolution might be a very favorable aspect of DG schemes because most engineering CED calculations are indeed done on resolution-starved meshes.

**Table VII shows the accuracy analysis for the second-order DGTD scheme for the propagation of an electromagnetic wave in vacuum. A CFL that was 95% of the maximum was used. The errors and accuracy in the y-component of the electric displacement vector (measured at the y-faces) and z-component of the magnetic induction (measured as zone averages) are shown.**

| Zones | Dy $L_1$ error | Dy $L_1$ accuracy | Dy $L_{inf}$ error | Dy $L_{inf}$ accuracy |
|---|---|---|---|---|
| $8^2$ | 2.66E-04 | | 3.85E-04 | |
| $16^2$ | 4.38E-05 | 2.60 | 6.74E-05 | 2.51 |
| $32^2$ | 7.30E-06 | 2.59 | 1.14E-05 | 2.57 |

| Zones | | | | |
|---|---|---|---|---|
| $64^2$ | 1.45E-06 | 2.33 | 2.28E-06 | 2.32 |
| $128^2$ | 3.33E-07 | 2.12 | 5.23E-07 | 2.12 |
| $256^2$ | 8.27E-08 | 2.01 | 1.30E-07 | 2.01 |
| $512^2$ | 2.05E-08 | 2.01 | 3.22E-08 | 2.01 |
| Zones | Bz $L_1$ error | Bz $L_1$ accuracy | Bz $L_{inf}$ error | Bz $L_{inf}$ accuracy |
| $8^2$ | 1.30E-01 | | 2.09E-01 | |
| $16^2$ | 2.24E-02 | 2.53 | 3.56E-02 | 2.55 |
| $32^2$ | 3.80E-03 | 2.56 | 5.98E-03 | 2.58 |
| $64^2$ | 7.61E-04 | 2.32 | 1.20E-03 | 2.32 |
| $128^2$ | 1.76E-04 | 2.11 | 2.77E-04 | 2.11 |
| $256^2$ | 4.39E-05 | 2.00 | 6.89E-05 | 2.00 |
| $512^2$ | 1.09E-05 | 2.01 | 1.71E-05 | 2.01 |

**Table VIII shows the accuracy analysis for the third-order DGTD scheme for the propagation of an electromagnetic wave in vacuum. A CFL that was 95% of the maximum was used. The errors and accuracy in the y-component of the electric displacement vector (measured at the y-faces) and z-component of the magnetic induction (measured as zone averages) are shown.**

| Zones | Dy $L_1$ error | Dy $L_1$ accuracy | Dy $L_{inf}$ error | Dy $L_{inf}$ accuracy |
|---|---|---|---|---|
| $8^2$ | 1.30E-04 | | 2.02E-04 | |
| $16^2$ | 1.45E-05 | 3.16 | 2.29E-05 | 3.14 |
| $32^2$ | 1.74E-06 | 3.06 | 2.74E-06 | 3.06 |
| $64^2$ | 2.14E-07 | 3.02 | 3.37E-07 | 3.02 |
| $128^2$ | 2.67E-08 | 3.01 | 4.19E-08 | 3.01 |
| $256^2$ | 3.32E-09 | 3.00 | 5.22E-09 | 3.00 |
| $512^2$ | 4.15E-10 | 3.00 | 6.52E-10 | 3.00 |

| Zones | Bz $L_1$ error | Bz $L_1$ accuracy | Bz $L_{inf}$ error | Bz $L_{inf}$ accuracy |
|---|---|---|---|---|
| $8^2$ | 6.55E-02 | | 9.52E-02 | |
| $16^2$ | 7.39E-03 | 3.15 | 1.13E-02 | 3.07 |
| $32^2$ | 9.03E-04 | 3.03 | 1.41E-03 | 3.01 |
| $64^2$ | 1.13E-04 | 3.00 | 1.77E-04 | 3.00 |
| $128^2$ | 1.41E-05 | 3.00 | 2.21E-05 | 3.00 |
| $256^2$ | 1.76E-06 | 3.00 | 2.77E-06 | 3.00 |
| $512^2$ | 2.21E-07 | 3.00 | 3.47E-07 | 3.00 |

**Table IX shows the accuracy analysis for the fourth-order DGTD scheme for the propagation of an electromagnetic wave in vacuum. A CFL that was 95% of the maximum was used. The errors and accuracy in the y-component of the electric displacement vector (measured at the y-faces) and z-component of the magnetic induction (measured as zone averages) are shown.**

| Zones | Dy $L_1$ error | Dy $L_1$ accuracy | Dy $L_{inf}$ error | Dy $L_{inf}$ accuracy |
|---|---|---|---|---|
| $8^2$ | 8.55E-06 | | 1.30E-05 | |
| $16^2$ | 5.97E-07 | 3.84 | 9.30E-07 | 3.80 |
| $32^2$ | 3.86E-08 | 3.95 | 6.04E-08 | 3.94 |
| $64^2$ | 2.43E-09 | 3.99 | 3.82E-09 | 3.98 |
| $128^2$ | 1.52E-10 | 4.00 | 2.39E-10 | 4.00 |
| $256^2$ | 9.54E-12 | 4.00 | 1.50E-11 | 4.00 |
| $512^2$ | 6.04E-13 | 3.98 | 9.49E-13 | 3.98 |
| Zones | Bz $L_1$ error | Bz $L_1$ accuracy | Bz $L_{inf}$ error | Bz $L_{inf}$ accuracy |
| $8^2$ | 3.65E-03 | | 5.62E-03 | |
| $16^2$ | 2.90E-04 | 3.65 | 4.54E-04 | 3.63 |
| $32^2$ | 1.97E-05 | 3.88 | 3.09E-05 | 3.88 |
| $64^2$ | 1.27E-06 | 3.96 | 2.00E-06 | 3.95 |
| $128^2$ | 8.04E-08 | 3.98 | 1.26E-07 | 3.98 |
| $256^2$ | 5.06E-09 | 3.99 | 7.94E-09 | 3.99 |
| $512^2$ | 3.21E-10 | 3.98 | 5.04E-10 | 3.98 |

It is interesting to ask how the results from Tables VII, VIII and IX change when a PNPM scheme is used. Table X shows a second order P0P1 scheme and should be compared to the second order DGTD results from Table VII. We see that both schemes are somewhat low accuracy but the DGTD results are superior to the P0P1 results. Tables XI and XII show results from third order P0P2 and P1P2 schemes and should be compared to the third order DGTD results from Table VIII. We see that the P0P2 is not as good as third order DGTD schemes. However, we see that P1P2 is entirely competitive with third order DGTD schemes. This underscores the fact that by retaining time-evolutionary equations for the linear variation, the P1PN schemes replicate most of the advantages of DGTD schemes for CED.

Tables XIII and XIV show the accuracy analysis for fourth order P0P3 and P1P3 schemes and they should be compared to the accuracy analysis for fourth order P3P3 schemes in Table IX. The P0P3 and P1P3 schemes have been run with a substantially larger timesteps than the P3P3 scheme. We clearly see that the accuracy of the P1P3 scheme in Table XIV is even slightly better than the accuracy of the P3P3 scheme. We attribute that to the substantially larger timesteps that can be taken by the P1P3 schemes.

**Table X shows the accuracy analysis for the second-order P0P1 scheme for the propagation of an electromagnetic wave in vacuum. A CFL that was 95% of the maximum was used. The errors and accuracy in the y-component of the electric displacement vector (measured at the y-faces) and z-component of the magnetic induction (measured as zone averages) are shown.**

| Zones | Dy $L_1$ error | Dy $L_1$ accuracy | Dy $L_{inf}$ error | Dy $L_{inf}$ accuracy |
|---|---|---|---|---|
| $8^2$ | 5.96E-04 | | 8.94E-04 | |
| $16^2$ | 2.14E-04 | 1.48 | 3.43E-04 | 1.38 |

| Zones | | | | |
|---|---|---|---|---|
| $32^2$ | 5.67E-05 | 1.91 | 1.01E-04 | 1.76 |
| $64^2$ | 1.47E-05 | 1.95 | 2.23E-05 | 2.18 |
| $128^2$ | 3.61E-06 | 2.02 | 5.66E-06 | 1.98 |
| $256^2$ | 9.02E-07 | 2.00 | 1.42E-06 | 2.00 |
| $512^2$ | 2.26E-07 | 2.00 | 3.54E-07 | 2.00 |
| Zones | Bz $L_1$ error | Bz $L_1$ accuracy | Bz $L_{inf}$ error | Bz $L_{inf}$ accuracy |
| $8^2$ | 3.17E-01 | | 4.92E-01 | |
| $16^2$ | 1.13E-01 | 1.49 | 1.77E-01 | 1.47 |
| $32^2$ | 3.02E-02 | 1.90 | 5.39E-02 | 1.72 |
| $64^2$ | 7.67E-03 | 1.98 | 1.19E-02 | 2.18 |
| $128^2$ | 1.92E-03 | 1.99 | 3.01E-03 | 1.99 |
| $256^2$ | 4.80E-04 | 2.00 | 7.55E-04 | 2.00 |
| $512^2$ | 1.20E-04 | 2.00 | 1.89E-04 | 2.00 |

**Table XI shows the accuracy analysis for the third-order P0P2 scheme for the propagation of an electromagnetic wave in vacuum. A CFL that was 95% of the maximum was used. The errors and accuracy in the y-component of the electric displacement vector (measured at the y-faces) and z-component of the magnetic induction (measured as zone averages) are shown.**

| Zones | Dy $L_1$ error | Dy $L_1$ accuracy | Dy $L_{inf}$ error | Dy $L_{inf}$ accuracy |
|---|---|---|---|---|
| $8^2$ | 3.37E-04 | | 5.16E-04 | |
| $16^2$ | 5.56E-05 | 2.60 | 8.20E-05 | 2.65 |
| $32^2$ | 1.06E-05 | 2.39 | 1.65E-05 | 2.31 |
| $64^2$ | 1.34E-06 | 2.98 | 2.11E-06 | 2.97 |
| $128^2$ | 1.68E-07 | 3.00 | 2.64E-07 | 2.99 |
| $256^2$ | 2.12E-08 | 2.99 | 3.33E-08 | 2.99 |
| $512^2$ | 2.70E-09 | 2.97 | 4.24E-09 | 2.97 |

| Zones | Bz $L_1$ error | Bz $L_1$ accuracy | Bz $L_{inf}$ error | Bz $L_{inf}$ accuracy |
|---|---|---|---|---|
| $8^2$ | 1.79E-01 | | 2.62E-01 | |
| $16^2$ | 2.97E-02 | 2.59 | 4.37E-02 | 2.58 |
| $32^2$ | 5.61E-03 | 2.40 | 8.81E-03 | 2.31 |
| $64^2$ | 7.15E-04 | 2.97 | 1.12E-03 | 2.97 |
| $128^2$ | 8.97E-05 | 2.99 | 1.41E-04 | 2.99 |
| $256^2$ | 1.13E-05 | 2.99 | 1.77E-05 | 2.99 |
| $512^2$ | 1.44E-06 | 2.97 | 2.26E-06 | 2.97 |

**Table XII shows the accuracy analysis for the third-order P1P2 scheme for the propagation of an electromagnetic wave in vacuum. A CFL that was 95% of the maximum was used. The errors and accuracy in the y-component of the electric displacement vector (measured at the y-faces) and z-component of the magnetic induction (measured as zone averages) are shown.**

| Zones | Dy $L_1$ error | Dy $L_1$ accuracy | Dy $L_{inf}$ error | Dy $L_{inf}$ accuracy |
|---|---|---|---|---|
| $8^2$ | 1.34E-04 | | 2.05E-04 | |
| $16^2$ | 1.94E-05 | 2.79 | 3.44E-05 | 2.57 |
| $32^2$ | 2.57E-06 | 2.92 | 4.99E-06 | 2.79 |
| $64^2$ | 2.79E-07 | 3.20 | 4.45E-07 | 3.49 |
| $128^2$ | 2.90E-08 | 3.27 | 4.49E-08 | 3.31 |
| $256^2$ | 3.61E-09 | 3.01 | 5.66E-09 | 2.99 |
| $512^2$ | 4.51E-10 | 3.00 | 7.09E-10 | 3.00 |
| Zones | Bz $L_1$ error | Bz $L_1$ accuracy | Bz $L_{inf}$ error | Bz $L_{inf}$ accuracy |
| $8^2$ | 6.81E-02 | | 9.98E-02 | |
| $16^2$ | 1.07E-02 | 2.67 | 1.92E-02 | 2.38 |
| $32^2$ | 1.40E-03 | 2.93 | 2.63E-03 | 2.87 |
| $64^2$ | 1.49E-04 | 3.23 | 2.35E-04 | 3.48 |
| $128^2$ | 1.55E-05 | 3.27 | 2.40E-05 | 3.29 |
| $256^2$ | 1.92E-06 | 3.01 | 3.02E-06 | 2.99 |
| $512^2$ | 2.41E-07 | 3.00 | 3.78E-07 | 3.00 |

Table XIII shows the accuracy analysis for the fourth-order P0P3 scheme for the propagation of an electromagnetic wave in vacuum. A CFL that was 95% of the maximum was used. The errors and accuracy in the y-component of the electric displacement vector (measured at the y-faces) and z-component of the magnetic induction (measured as zone averages) are shown.

| Zones | Dy $L_1$ error | Dy $L_1$ accuracy | Dy $L_{inf}$ error | Dy $L_{inf}$ accuracy |
|---|---|---|---|---|
| $8^2$ | 7.73E-05 | | 1.22E-04 | |
| $16^2$ | 2.60E-06 | 4.90 | 4.02E-06 | 4.93 |
| $32^2$ | 8.47E-08 | 4.94 | 1.56E-07 | 4.69 |
| $64^2$ | 2.97E-08 | 1.51 | 4.65E-08 | 1.74 |
| $128^2$ | 1.95E-09 | 3.92 | 3.07E-09 | 3.92 |
| $256^2$ | 1.52E-10 | 3.69 | 2.38E-10 | 3.69 |
| $512^2$ | 9.51E-12 | 4.00 | 1.49E-11 | 4.00 |
| Zones | Bz $L_1$ error | Bz $L_1$ accuracy | Bz $L_{inf}$ error | Bz $L_{inf}$ accuracy |
| $8^2$ | 4.01E-02 | | 6.42E-02 | |
| $16^2$ | 1.36E-03 | 4.88 | 2.14E-03 | 4.90 |
| $32^2$ | 4.52E-05 | 4.91 | 8.10E-05 | 4.72 |
| $64^2$ | 1.58E-05 | 1.51 | 2.48E-05 | 1.71 |
| $128^2$ | 1.04E-06 | 3.92 | 1.63E-06 | 3.92 |
| $256^2$ | 8.04E-08 | 3.69 | 1.27E-07 | 3.69 |
| $512^2$ | 5.07E-09 | 4.00 | 7.96E-09 | 4.00 |

Table XIV shows the accuracy analysis for the fourth-order P1P3 scheme for the propagation of an electromagnetic wave in vacuum. A CFL that was 95% of the maximum was used. The errors and **accuracy in the y-component of the electric displacement vector (measured at the y-faces) and z-component of the magnetic induction (measured as zone averages) are shown.**

| Zones | Dy $L_1$ error | Dy $L_1$ accuracy | Dy $L_{inf}$ error | Dy $L_{inf}$ accuracy |
|---|---|---|---|---|
| $8^2$ | 3.43E-05 | | 5.43E-05 | |
| $16^2$ | 1.36E-06 | 4.66 | 2.28E-06 | 4.57 |
| $32^2$ | 5.11E-08 | 4.74 | 8.07E-08 | 4.82 |
| $64^2$ | 2.07E-09 | 4.62 | 3.26E-09 | 4.63 |
| $128^2$ | 1.00E-10 | 4.37 | 1.58E-10 | 4.37 |
| $256^2$ | 5.81E-12 | 4.11 | 9.13E-12 | 4.11 |
| $512^2$ | 3.51E-12 | 4.05 | 5.52E-13 | 4.05 |
| Zones | Bz $L_1$ error | Bz $L_1$ accuracy | Bz $L_{inf}$ error | Bz $L_{inf}$ accuracy |
| $8^2$ | 1.88E-02 | | 2.84E-02 | |
| $16^2$ | 7.40E-04 | 4.66 | 1.23E-03 | 4.53 |
| $32^2$ | 2.77E-05 | 4.74 | 4.35E-05 | 4.82 |
| $64^2$ | 1.12E-06 | 4.63 | 1.76E-06 | 4.63 |
| $128^2$ | 5.40E-08 | 4.37 | 8.48E-08 | 4.38 |
| $256^2$ | 3.11E-09 | 4.12 | 4.89E-09 | 4.12 |
| $512^2$ | 1.88E-10 | 4.05 | 2.95E-10 | 4.05 |

While Maxwell's equations (for non-dispersive media) constitute a linear system, the differential form of Maxwell's equations also ensures the conservation of electromagnetic energy. The electromagnetic energy is quadratic in terms of the electromagnetic field. When the conductivity is zero, the time rate of change of the electromagnetic energy density is given by the divergence of the Poynting flux. Our governing equations are linear, with the result that we do not do anything special to ensure quadratic energy conservation. Nevertheless, we expect that if the numerical scheme is consistent and highly accurate it should conserve the electromagnetic energy on the computational mesh with almost perfect precision. Of course, when an electromagnetic wave spans a large number of zones, any numerical method will conserve the electromagnetic energy on the computational mesh with near-perfect precision. The measure of excellence is, therefore, that the quadratic electromagnetic energy is almost perfectly conserved with the smallest number of zones per wavelength.

The data that we have extracted for this test problem makes it easy for us to quantify how well each of the reported schemes conserves electromagnetic energy. Fig 9a shows the electromagnetic energy after one periodic orbit as a function of number of zones along one direction of the two-dimensional mesh for P=1, P=2 and P=3 DG schemes. Fig. 9b shows the same information for P0P1, P0P2 and P0P3 schemes. Fig. 9c shows the same information for P1P2 and P1P3 schemes. From Fig. 9a it is easy to see that the second order P=1 DGTD scheme does not conserve electromagnetic energy till we reach a mesh with 32 zones in each direction. In contrast, Fig. 9a also shows that the fourth order P=3 DGTD scheme does an excellent job of energy conservation even when we have a mesh with 16 zones in each direction; actually even with 8 zones in each direction, it does quite well. We also see from Fig. 9a that the third order P=2 DGTD scheme has an intermediate performance between the second and fourth order DGTD schemes. The P0PM schemes in Fig. 9b are essentially the second, third and fourth order accurate FVTD schemes from Balsara *et al*. [19], [20]. Comparing Fig. 9b and Fig. 9a we see that on resolution-starved meshes the P0PM schemes are inferior performers when it comes to energy conservation, though on somewhat larger meshes they do begin to perform very well. Fig. 9c shows the ability of third order P1P2 and fourth order P1P3 schemes to conserve electromagnetic energy. Comparing Fig. 9c to Fig. 9a, we see that the P1PM schemes conserve energy almost as well as the DGTD schemes. This again underscores the fact that evolving the first moments according to the governing equations described in this paper makes a huge impact on the quality of the simulation. Fig. 9 helps firm up our viewpoint that P1PM schemes offer almost the same advantages as DFTD schemes; however the P1PM schemes operate with significantly larger CFL numbers.

## VII) Conclusions

This paper represents the first time that globally constraint-preserving DGTD schemes have been designed for CED. The algorithms presented here are based on a novel DG-like method that is applied to a Yee-type staggering of the electromagnetic field variables in the faces of the mesh. The other two novel building blocks of the method include constraint-preserving reconstruction of the electromagnetic fields and multidimensional Riemann solvers; both of which have been developed in recent years by the first author. While we have explicitly catalogued DGTD schemes for CED up to fourth order in accuracy, there is in principle no barrier to going to even higher order. DG schemes have another attractive feature for engineering CED which is that they take well to general geometries.

Since the schemes are linear, it is possible to carry out a von Neumann stability analysis of the entire suite of DGTD and PNPM schemes for CED at orders of accuracy ranging from second to fourth. The analysis requires some simplifications in order to make it analytically tractable, however, it proves to be extremely instructive. Our stability analysis gives us the maximal CFL numbers that can be sustained by the DGTD and PNPM schemes presented here at all orders (upto four). It also enables us to understand the wave propagation characteristics of the schemes in various directions on a Cartesian mesh. We find that constraint-preserving DGTD schemes permit CFL numbers that are competitive with conventional DG schemes. However, the permissible CFL decreases with increasing order of accuracy for the DGTD schemes. The P1PM schemes counteract this trend, offering performance that is competitive with DGTD schemes of the same order, while also supporting a robust CFL. We also find that the third and fourth order constraint-preserving DGTD schemes have some extremely attractive properties when it comes to low-dispersion, low-dissipation propagation of electromagnetic waves in multidimensions. Numerical accuracy tests are also provided to support the von Neumann stability analysis. The higher order members of the family of schemes presented here also do an excellent job on conservation of electromagnetic energy. We expect these methods to play a role in those problems of engineering CED where exceptional precision must be achieved at any cost.

**Acknowledgements**

DSB acknowledges support via NSF grants NSF-ACI-1533850, NSF-DMS-1622457 and NSF_ACI-1713765. Several simulations were performed on a cluster at UND that is run by the Center for Research Computing. Computer support on NSF's XSEDE and Blue Waters computing resources is also acknowledged. RK acknowledges support by the Swiss National Science

Foundation (SNSF) under grant 200021-169631. R.K. also acknowledges the use of computational resources provided by the Swiss SuperComputing Center (CSCS), under the allocation grant s661, s665, s667 and s744. The authors acknowledge the computational resources provided by the EULER cluster of ETHZ.

**Appendix A**

In this appendix we explicitly document the 36 coefficients of the matrix in eqn. (3.16). This enables the reader to cross-check his or her mathematics. These coefficients are easily obtained with the help of a computer algebra system. For the first row of the matrix, we have

$$A_{11} = \frac{ce^{-ik_y\Delta y}\left(e^{ik_y\Delta y}-1\right)^2}{2\Delta y} \quad ; \quad A_{12} = -\frac{ce^{-ik_y\Delta y}\left(e^{ik_y\Delta y}-1\right)\left(e^{ik_y\Delta y}+1\right)}{4\Delta y} \quad ;$$

$$A_{13} = \frac{c\left(e^{ik_x\Delta x}+1\right)\left(e^{ik_y\Delta y}-1\right)e^{-i(k_x\Delta x+k_y\Delta y)/2}}{4\Delta y} \quad ; \quad A_{14} = \frac{c^2\left(e^{ik_x\Delta x}+1\right)\left(e^{ik_y\Delta y}-1\right)\left(e^{ik_y\Delta y}+1\right)e^{-i(k_x\Delta x+2k_y\Delta y)/2}}{4\Delta y} \quad ;$$

$$A_{15} = -\frac{c^2\left(e^{ik_x\Delta x}-1\right)\left(e^{ik_y\Delta y}-1\right)\left(e^{ik_y\Delta y}+1\right)e^{-i(k_x\Delta x+2k_y\Delta y)/2}}{8\Delta y} \quad ;$$

$$A_{16} = -\frac{c^2\left(e^{ik_x\Delta x}+1\right)\left(e^{ik_y\Delta y}-1\right)^2 e^{-i(k_x\Delta x+2k_y\Delta y)/2}}{8\Delta y} \quad ;$$

For the second row of the matrix, we have

$$A_{21} = \frac{3ce^{-ik_y\Delta y}\left(e^{ik_y\Delta y}-1\right)\left(e^{ik_y\Delta y}+1\right)}{\Delta y} \quad ;$$

$$A_{22} = -\frac{ce^{-ik_y\Delta y-ik_x\Delta x}\left(3e^{2ik_y\Delta y+ik_x\Delta x}-e^{ik_y\Delta y+2ik_x\Delta x}+8e^{ik_y\Delta y+ik_x\Delta x}-e^{ik_y\Delta y}+3e^{ik_x\Delta x}\right)}{2\Delta y} \quad ; \quad A_{23}=0 \quad ;$$

$$A_{24} = \frac{3c^2\left(e^{ik_x\Delta x}+1\right)\left(e^{ik_y\Delta y}-1\right)^2 e^{-i(k_x\Delta x+2k_y\Delta y)/2}}{2\Delta y} \quad ; \quad A_{25} = -\frac{3c^2\left(e^{ik_x\Delta x}-1\right)\left(e^{ik_y\Delta y}-1\right)^2 e^{-i(k_x\Delta x+2k_y\Delta y)/2}}{4\Delta y} \quad ;$$

$$A_{26} = -\frac{3c^2\left(e^{ik_x\Delta x}+1\right)\left(e^{ik_y\Delta y}-1\right)\left(e^{ik_y\Delta y}+1\right)e^{-i(k_x\Delta x+2k_y\Delta y)/2}}{4\Delta y} \quad ;$$

For the third row of the matrix, we have

$$A_{31}=0 \quad ; \quad A_{32}=0 \quad ; \quad A_{33} = \frac{ce^{-ik_y\Delta y-ik_x\Delta x}\left(e^{2ik_y\Delta y+ik_x\Delta x}-3e^{ik_y\Delta y+2ik_x\Delta x}-8e^{ik_y\Delta y+ik_x\Delta x}-3e^{ik_y\Delta y}+e^{ik_x\Delta x}\right)}{2\Delta x} \quad ;$$

$$A_{34} = -\frac{3c^2\left(e^{ik_x\Delta x}-1\right)^2\left(e^{ik_y\Delta y}+1\right)e^{-i(k_x\Delta x+2k_y\Delta y)/2}}{2\Delta x} \quad ;$$

$$A_{35} = \frac{3c^2\left(e^{ik_x\Delta x}-1\right)\left(e^{ik_x\Delta x}+1\right)\left(e^{ik_y\Delta y}+1\right)e^{-i(k_x\Delta x+2k_y\Delta y)/2}}{4\Delta x} \quad;$$

$$A_{36} = \frac{3c^2\left(e^{ik_x\Delta x}-1\right)^2\left(e^{ik_y\Delta y}-1\right)e^{-i(k_x\Delta x+2k_y\Delta y)/2}}{4\Delta x} \quad;$$

For the fourth row of the matrix, we have

$$A_{41} = \frac{\left(e^{ik_x\Delta x}+1\right)\left(e^{ik_y\Delta y}-1\right)\left(e^{ik_y\Delta y}+1\right)e^{-i(k_x\Delta x+2k_y\Delta y)/2}}{4\Delta y} \quad;$$

$$A_{42} = -\frac{\left(e^{ik_x\Delta x}+1\right)e^{-ik_y\Delta y-\frac{3ik_x\Delta x}{2}}}{24\Delta x\Delta y}(3\Delta x e^{2ik_y\Delta y+ik_x\Delta x}+\Delta y e^{ik_y\Delta y+2ik_x\Delta x}-2\Delta y e^{ik_y\Delta y+ik_x\Delta x}-6\Delta x e^{ik_y\Delta y+ik_x\Delta x}$$
$$+\Delta y e^{ik_y\Delta y}+3\Delta x e^{ik_x\Delta x})$$

$$A_{43} = \frac{\left(e^{ik_y\Delta y}+1\right)e^{-\frac{3ik_y\Delta y}{2}-ik_x\Delta x}}{24\Delta x\Delta y}(\Delta x e^{2ik_y\Delta y+ik_x\Delta x}+3\Delta y e^{ik_y\Delta y+2ik_x\Delta x}-6\Delta y e^{ik_y\Delta y+ik_x\Delta x}-2\Delta x e^{ik_y\Delta y+ik_x\Delta x}$$
$$+3\Delta y e^{ik_y\Delta y}+\Delta x e^{ik_x\Delta x})$$

$$A_{44} = \frac{ce^{-ik_y\Delta y-ik_x\Delta x}}{2\Delta x\Delta y}(\Delta x e^{2ik_y\Delta y+ik_x\Delta x}+\Delta y e^{ik_y\Delta y+2ik_x\Delta x}-2\Delta y e^{ik_y\Delta y+ik_x\Delta x}-2\Delta x e^{ik_y\Delta y+ik_x\Delta x}+\Delta y e^{ik_y\Delta y}+\Delta x e^{ik_x\Delta x})$$

$$A_{45} = -\frac{ce^{-ik_x\Delta x}\left(e^{ik_x\Delta x}-1\right)\left(e^{ik_x\Delta x}+1\right)}{4\Delta x} \quad; \quad A_{46} = -\frac{ce^{-ik_y\Delta y}\left(e^{ik_y\Delta y}-1\right)\left(e^{ik_y\Delta y}+1\right)}{4\Delta y} \quad;$$

For the fifth row of the matrix, we have

$$A_{51} = \frac{\left(e^{ik_x\Delta x}-1\right)\left(e^{ik_y\Delta y}-1\right)\left(e^{ik_y\Delta y}+1\right)e^{-i(k_x\Delta x+2k_y\Delta y)/2}}{2\Delta y} \quad;$$

$$A_{52} = -\frac{\left(e^{ik_x\Delta x}-1\right)e^{-ik_y\Delta y-\frac{3ik_x\Delta x}{2}}}{4\Delta x\Delta y}(\Delta x e^{2ik_y\Delta y+ik_x\Delta x}+\Delta y e^{ik_y\Delta y+2ik_x\Delta x}-2\Delta y e^{ik_y\Delta y+ik_x\Delta x}-2\Delta x e^{ik_y\Delta y+ik_x\Delta x}$$
$$+\Delta y e^{ik_y\Delta y}+\Delta x e^{ik_x\Delta x})$$

$$A_{53} = \frac{3\left(e^{ik_x\Delta x}-1\right)\left(e^{ik_x\Delta x}+1\right)\left(e^{ik_y\Delta y}+1\right)e^{-i(2k_x\Delta x+k_y\Delta y)/2}}{4\Delta x} \quad ; \quad A_{54} = \frac{3ce^{-ik_x\Delta x}\left(e^{ik_x\Delta x}-1\right)\left(e^{ik_x\Delta x}+1\right)}{\Delta x} \quad ;$$

$$A_{55} = \frac{ce^{-ik_y\Delta y-ik_x\Delta x}}{2\Delta x\Delta y}(\Delta x e^{2ik_y\Delta y+ik_x\Delta x}-3\Delta y e^{ik_y\Delta y+2ik_x\Delta x}-6\Delta y e^{ik_y\Delta y+ik_x\Delta x}-2\Delta x e^{ik_y\Delta y+ik_x\Delta x}-3\Delta y e^{ik_y\Delta y}$$
$$+\Delta x e^{ik_x\Delta x})$$

$$A_{56} = 0$$

For the sixth row of the matrix, we have

$$A_{61} = \frac{3\left(e^{ik_x\Delta x}+1\right)\left(e^{ik_y\Delta y}-1\right)^2 e^{-i(k_x\Delta x+2k_y\Delta y)/2}}{2\Delta y} \quad ;$$

$$A_{62} = -\frac{3\left(e^{ik_x\Delta x}+1\right)\left(e^{ik_y\Delta y}-1\right)\left(e^{ik_y\Delta y}+1\right)e^{-i(k_x\Delta x+2k_y\Delta y)/2}}{4\Delta y} \quad ;$$

$$A_{63} = \frac{\left(e^{ik_y\Delta y}-1\right)e^{-\frac{3ik_y\Delta y}{2}-ik_x\Delta x}}{4\Delta x\Delta y}(\Delta x e^{2ik_y\Delta y+ik_x\Delta x}+\Delta y e^{ik_y\Delta y+2ik_x\Delta x}-2\Delta y e^{ik_y\Delta y+ik_x\Delta x}-2\Delta x e^{ik_y\Delta y+ik_x\Delta x}$$
$$+\Delta y e^{ik_y\Delta y}+\Delta x e^{ik_x\Delta x})$$

$$A_{64} = \frac{3ce^{-ik_y\Delta y}\left(e^{ik_y\Delta y}-1\right)\left(e^{ik_y\Delta y}+1\right)}{\Delta y} \quad ; \quad A_{65} = 0$$

$$A_{66} = -\frac{ce^{-ik_y\Delta y-ik_x\Delta x}}{2\Delta x\Delta y}(3\Delta x e^{2ik_y\Delta y+ik_x\Delta x}-\Delta y e^{ik_y\Delta y+2ik_x\Delta x}+2\Delta y e^{ik_y\Delta y+ik_x\Delta x}+6\Delta x e^{ik_y\Delta y+ik_x\Delta x}$$
$$-\Delta y e^{ik_y\Delta y}+3\Delta x e^{ik_x\Delta x})$$

This completes our catalogue of all the 36 matrix elements of the matrix "**A**" in eqn. (3.16).

**Appendix B**

The analogue of eqn. (4.1) for the facial *x*-component of the electric displacement at fourth order can be written as

$$D^x(y,t) = D_0^x(t) + D_y^x(t)\left(\frac{y}{\Delta y}\right) + D_{yy}^x(t)\left[\left(\frac{y}{\Delta y}\right)^2 - \frac{1}{12}\right] + D_{yyy}^x(t)\left[\left(\frac{y}{\Delta y}\right)^3 - \frac{3}{20}\left(\frac{y}{\Delta y}\right)\right] \quad \text{(B.1)}$$

Because we have used orthogonal basis sets, the evolutionary equations for $D_0^x(t)$, $D_y^x(t)$ and $D_{yy}^x(t)$ are still given by eqns. (3.4), (3.5) and (4.2) respectively. The analogue of eqn. (4.3) for the facial *y*-component of the electric displacement at fourth order can be written as

$$D^y(x,t) = D_0^y(t) + D_x^y(t)\left(\frac{x}{\Delta x}\right) + D_{xx}^y(t)\left[\left(\frac{x}{\Delta x}\right)^2 - \frac{1}{12}\right] + D_{xxx}^y(t)\left[\left(\frac{x}{\Delta x}\right)^3 - \frac{3}{20}\left(\frac{x}{\Delta x}\right)\right] \quad \text{(B.2)}$$

Because we have used orthogonal basis sets, the evolutionary equations for $D_0^y(t)$, $D_x^y(t)$ and $D_{xx}^y(t)$ are still given by (3.7), (3.8) and (4.4) respectively.

At fourth order, we can make a constraint-preserving reconstruction of the electric displacement that matches the variation of the electric displacement at the boundaries of the zone. Such a reconstruction has been described in Appendix B of Balsara *et al*. [20] and we repeat just a little bit of it because it helps us to explain the essential idea that changes for a DGTD scheme. The *x*-component of the reconstructed electric displacement within the zone is given by

$$D^x(x,y) = a_0 + a_x\left(\frac{x}{\Delta x}\right) + a_y\left(\frac{y}{\Delta y}\right) + a_{xx}\left(\left(\frac{x}{\Delta x}\right)^2 - \frac{1}{12}\right) + a_{yy}\left(\left(\frac{y}{\Delta y}\right)^2 - \frac{1}{12}\right) + a_{xy}\left(\frac{x}{\Delta x}\right)\left(\frac{y}{\Delta y}\right)$$

$$+ a_{xxx}\left(\left(\frac{x}{\Delta x}\right)^3 - \frac{3}{20}\left(\frac{x}{\Delta x}\right)\right) + a_{yyy}\left(\left(\frac{y}{\Delta y}\right)^3 - \frac{3}{20}\left(\frac{y}{\Delta y}\right)\right) + a_{xxy}\left(\left(\frac{x}{\Delta x}\right)^2 - \frac{1}{12}\right)\left(\frac{y}{\Delta y}\right) + a_{xyy}\left(\frac{x}{\Delta x}\right)\left(\left(\frac{y}{\Delta y}\right)^2 - \frac{1}{12}\right)$$

$$+ a_{xxxx}\left(\left(\frac{x}{\Delta x}\right)^4 - \frac{3}{14}\left(\frac{x}{\Delta x}\right)^2 + \frac{3}{560}\right) + a_{xyyy}\left(\frac{x}{\Delta x}\right)\left(\left(\frac{y}{\Delta y}\right)^3 - \frac{3}{20}\left(\frac{y}{\Delta y}\right)\right) + a_{xxxy}\left(\left(\frac{x}{\Delta x}\right)^3 - \frac{3}{20}\left(\frac{x}{\Delta x}\right)\right)\left(\frac{y}{\Delta y}\right)$$

$$+ a_{xxyy}\left(\left(\frac{x}{\Delta x}\right)^2 - \frac{1}{12}\right)\left(\left(\frac{y}{\Delta y}\right)^2 - \frac{1}{12}\right) + a_{xxxxy}\left(\left(\frac{x}{\Delta x}\right)^4 - \frac{3}{14}\left(\frac{x}{\Delta x}\right)^2 + \frac{3}{560}\right)\left(\frac{y}{\Delta y}\right)$$

$$+ a_{xxyyy}\left(\left(\frac{x}{\Delta x}\right)^2 - \frac{1}{12}\right)\left(\left(\frac{y}{\Delta y}\right)^3 - \frac{3}{20}\left(\frac{y}{\Delta y}\right)\right)$$

(B.3)

The *y*-component of the reconstructed electric displacement within the zone is given by

$$D^y(x,y) = b_0 + b_x\left(\frac{x}{\Delta x}\right) + b_y\left(\frac{y}{\Delta y}\right) + b_{xx}\left(\left(\frac{x}{\Delta x}\right)^2 - \frac{1}{12}\right) + b_{yy}\left(\left(\frac{y}{\Delta y}\right)^2 - \frac{1}{12}\right) + b_{xy}\left(\frac{x}{\Delta x}\right)\left(\frac{y}{\Delta y}\right)$$

$$+ b_{xxx}\left(\left(\frac{x}{\Delta x}\right)^3 - \frac{3}{20}\left(\frac{x}{\Delta x}\right)\right) + b_{yyy}\left(\left(\frac{y}{\Delta y}\right)^3 - \frac{3}{20}\left(\frac{y}{\Delta y}\right)\right) + b_{xxy}\left(\left(\frac{x}{\Delta x}\right)^2 - \frac{1}{12}\right)\left(\frac{y}{\Delta y}\right) + b_{xyy}\left(\frac{x}{\Delta x}\right)\left(\left(\frac{y}{\Delta y}\right)^2 - \frac{1}{12}\right)$$

$$+ b_{yyyy}\left(\left(\frac{y}{\Delta y}\right)^4 - \frac{3}{14}\left(\frac{y}{\Delta y}\right)^2 + \frac{3}{560}\right) + b_{xxxy}\left(\left(\frac{x}{\Delta x}\right)^3 - \frac{3}{20}\left(\frac{x}{\Delta x}\right)\right)\left(\frac{y}{\Delta y}\right) + b_{xxyy}\left(\left(\frac{x}{\Delta x}\right)^2 - \frac{1}{12}\right)\left(\left(\frac{y}{\Delta y}\right)^2 - \frac{1}{12}\right)$$

$$+ b_{xyyy}\left(\frac{x}{\Delta x}\right)\left(\left(\frac{y}{\Delta y}\right)^3 - \frac{3}{20}\left(\frac{y}{\Delta y}\right)\right) + b_{xyyyy}\left(\frac{x}{\Delta x}\right)\left(\left(\frac{y}{\Delta y}\right)^4 - \frac{3}{14}\left(\frac{y}{\Delta y}\right)^2 + \frac{3}{560}\right)$$

$$+ b_{xxxyy}\left(\left(\frac{x}{\Delta x}\right)^3 - \frac{3}{20}\left(\frac{x}{\Delta x}\right)\right)\left(\left(\frac{y}{\Delta y}\right)^2 - \frac{1}{12}\right)$$

(B.4)

All the coefficients, except two, in the above two equations are set by the cubic variation within the faces of the mesh. The only two exceptions are the $a_{xxy}$ and $b_{xyy}$ terms in eqns. (B.3) and (B.4). For the FVDT scheme that was presented in Balsara *et al.* [20], these terms can be set via a WENO reconstruction process. For the DGTD scheme presented here, these terms are evolved via a conventional, zone-centered DG scheme. This is the important point of difference that we wanted to bring out for DGTD schemes at fourth and higher orders.

The analogue of eqn. (4.3) for the $z$-component of the magnetic induction at third order can be written as

$$B^z(x,y,t) = B_0^z(t) + B_x^z(t)\left(\frac{x}{\Delta x}\right) + B_y^z(t)\left(\frac{y}{\Delta y}\right) + B_{xx}^z(t)\left(\left(\frac{x}{\Delta x}\right)^2 - \frac{1}{12}\right) + B_{yy}^z(t)\left(\left(\frac{y}{\Delta y}\right)^2 - \frac{1}{12}\right)$$
$$+ B_{xy}^z(t)\left(\frac{x}{\Delta x}\right)\left(\frac{y}{\Delta y}\right) + B_{xxx}^z(t)\left(\left(\frac{x}{\Delta x}\right)^3 - \frac{3}{20}\left(\frac{x}{\Delta x}\right)\right) + B_{yyy}^z(t)\left(\left(\frac{y}{\Delta y}\right)^3 - \frac{3}{20}\left(\frac{y}{\Delta y}\right)\right)$$
$$+ B_{xxy}^z(t)\left(\left(\frac{x}{\Delta x}\right)^2 - \frac{1}{12}\right)\left(\frac{y}{\Delta y}\right) + B_{xyy}^z(t)\left(\frac{x}{\Delta x}\right)\left(\left(\frac{y}{\Delta y}\right)^2 - \frac{1}{12}\right)$$

(B.5)

Because we have used orthogonal basis sets, the evolutionary equations for $B_0^z(t)$, $B_x^z(t)$, $B_y^z(t)$, $B_{xx}^z(t)$, $B_{yy}^z(t)$ and $B_{xy}^z(t)$ are given by eqns. (3.11), (3.12), (3.13), (4.6), (4.7) and (4.8) respectively.

Let us first focus on obtaining the remaining update equations for the electric displacement. The update equation for $D_{yyy}^x(t)$ is obtained by using $\hat{\mathbf{n}} = \hat{\mathbf{x}}$ and the test function $\phi(y) = \left((y/\Delta y)^3 - 3(y/\Delta y)/20\right)$ in eqn. (3.2) to get

$$\frac{1}{2800}\frac{dD_{yyy}^x(t)}{dt} - \frac{1}{\mu}\frac{1}{20\Delta y}\left[B^{z**}(x = \Delta x/2, y = \Delta y/2) + B^{z**}(x = \Delta x/2, y = -\Delta y/2)\right]$$
$$+ \frac{1}{\mu}\frac{3}{\Delta y}\left\langle\left((y/\Delta y)^2 - 1/20\right)B^{z*}(x = \Delta x/2, y)\right\rangle = 0$$

(B.6)

The update equation for $D_{xxx}^y(t)$ is obtained by using $\hat{\mathbf{n}} = \hat{\mathbf{y}}$ and the test function $\phi(x) = \left((x/\Delta x)^3 - 3(x/\Delta x)/20\right)$ in eqn. (3.2) to get

$$\frac{1}{2800}\frac{dD_{xxx}^{y}(t)}{dt} + \frac{1}{\mu}\frac{1}{20\Delta x}\left[B^{z**}\left(x=\Delta x/2, y=\Delta y/2\right) + B^{z**}\left(x=-\Delta x/2, y=\Delta y/2\right)\right]$$
$$-\frac{1}{\mu}\frac{3}{\Delta x}\left\langle\left(\left(x/\Delta x\right)^{2}-1/20\right)B^{z*}\left(x, y=\Delta y/2\right)\right\rangle = 0$$
(B.7)

Realize too that we have chosen to endow two of the modes in eqns. (B.3) and (B.4) with time-dependence. The orthogonality of the modes in eqns. (B.3) and (B.4) proves to be very helpful in what follows. We can, therefore, write those terms as $a_{xxy}(t)$ and $b_{xyy}(t)$. The update equation for $a_{xxy}(t)$ is obtained by applying a classical DG formulation with a test function $\phi(x, y) = \left(\left(x/\Delta x\right)^{2} - 1/12\right)\left(y/\Delta y\right)$ to the first row of eqn. (2.5) to get

$$\frac{1}{2160}\frac{da_{xxy}(t)}{dt} - \frac{1}{\mu}\frac{1}{2\Delta y}\left[\left\langle\left(\left(x/\Delta x\right)^{2}-1/12\right)B^{z*}\left(x, y=\Delta y/2\right)\right\rangle + \left\langle\left(\left(x/\Delta x\right)^{2}-1/12\right)B^{z*}\left(x, y=-\Delta y/2\right)\right\rangle\right]$$
$$+\frac{1}{\mu}\frac{1}{\Delta y}\left\{\left(\left(x/\Delta x\right)^{2}-1/12\right)B^{z}\left(x, y\right)\right\} = 0$$

(B.8)

The update equation for $b_{xyy}(t)$ is obtained by applying a classical DG formulation with a test function $\phi(x, y) = \left(x/\Delta x\right)\left(\left(y/\Delta y\right)^{2} - 1/12\right)$ to the second row of eqn. (2.5) to get

$$\frac{1}{2160}\frac{db_{xyy}(t)}{dt} + \frac{1}{\mu}\frac{1}{2\Delta x}\left[\left\langle\left(\left(y/\Delta y\right)^{2}-1/12\right)B^{z*}\left(x=\Delta x/2, y\right)\right\rangle + \left\langle\left(\left(y/\Delta y\right)^{2}-1/12\right)B^{z*}\left(x=-\Delta x/2, y\right)\right\rangle\right]$$
$$-\frac{1}{\mu}\frac{1}{\Delta x}\left\{\left(\left(y/\Delta y\right)^{2}-1/12\right)B^{z}\left(x, y\right)\right\} = 0$$

(B.9)

It is important to realize that at fourth order the DGTD update relies on the two facial updates in eqns. (B.6) and (B.7) as well as the updates for the volumetric terms in eqns. (B.8) and (B.9).

The update equation for $B_{xxx}^{z}(t)$ is obtained by using $\hat{\mathbf{n}} = \hat{\mathbf{z}}$ and the test function $\phi(x, y) = \left(\left(x/\Delta x\right)^{3} - 3\left(x/\Delta x\right)/20\right)$ in eqn. (3.9) to get

$$\frac{1}{2800}\frac{dB^z_{xxx}(t)}{dt}+\frac{1}{\varepsilon}\frac{1}{20\Delta x}\left[\left\langle D^{y*}(x=\Delta x/2,y)\right\rangle+\left\langle D^{y*}(x=-\Delta x/2,y)\right\rangle\right]$$

$$-\frac{1}{\varepsilon}\frac{1}{\Delta y}\left[\begin{array}{l}\left\langle\left((x/\Delta x)^3-3(x/\Delta x)/20\right)D^{x*}(x,y=\Delta y/2)\right\rangle\\-\left\langle\left((x/\Delta x)^3-3(x/\Delta x)/20\right)D^{x*}(x,y=-\Delta y/2)\right\rangle\end{array}\right]-\frac{1}{\varepsilon}\frac{3}{\Delta x}\left\{\left((x/\Delta x)^2-1/20\right)D^y(x,y)\right\}=0$$

(B.10)

The update equation for $B^z_{yyy}(t)$ is obtained by using $\hat{\mathbf{n}}=\hat{\mathbf{z}}$ and the test function $\phi(x,y)=\left((y/\Delta y)^3-3(y/\Delta y)/20\right)$ in eqn. (3.9) to get

$$\frac{1}{2800}\frac{dB^z_{yyy}(t)}{dt}+\frac{1}{\varepsilon}\frac{1}{\Delta x}\left[\begin{array}{l}\left\langle\left((y/\Delta y)^3-3(y/\Delta y)/20\right)D^{y*}(x=\Delta x/2,y)\right\rangle\\-\left\langle\left((y/\Delta y)^3-3(y/\Delta y)/20\right)D^{y*}(x=-\Delta x/2,y)\right\rangle\end{array}\right]$$

$$-\frac{1}{\varepsilon}\frac{1}{20\Delta y}\left[\left\langle D^{x*}(x,y=\Delta y/2)\right\rangle+\left\langle D^{x*}(x,y=-\Delta y/2)\right\rangle\right]+\frac{1}{\varepsilon}\frac{3}{\Delta y}\left\{\left((y/\Delta y)^2-1/20\right)D^x(x,y)\right\}=0$$

(B.11)

The update equation for $B^z_{xxy}(t)$ is obtained by using $\hat{\mathbf{n}}=\hat{\mathbf{z}}$ and the test function $\phi(x,y)=\left((x/\Delta x)^2-1/12\right)(y/\Delta y)$ in eqn. (3.9) to get

$$\frac{1}{2160}\frac{dB^z_{xxy}(t)}{dt}+\frac{1}{\varepsilon}\frac{1}{6\Delta x}\left[\left\langle(y/\Delta y)D^{y*}(x=\Delta x/2,y)\right\rangle-\left\langle(y/\Delta y)D^{y*}(x=-\Delta x/2,y)\right\rangle\right]$$

$$-\frac{1}{\varepsilon}\frac{1}{2\Delta y}\left[\left\langle\left((x/\Delta x)^2-1/12\right)D^{x*}(x,y=\Delta y/2)\right\rangle+\left\langle\left((x/\Delta x)^2-1/12\right)D^{x*}(x,y=-\Delta y/2)\right\rangle\right]\quad\text{(B.12)}$$

$$-\frac{1}{\varepsilon}\frac{2}{\Delta x}\left\{(x/\Delta x)(y/\Delta y)D^y(x,y)\right\}+\frac{1}{\varepsilon}\frac{1}{\Delta y}\left\{\left((x/\Delta x)^2-1/12\right)D^x(x,y)\right\}=0$$

The update equation for $B^z_{xyy}(t)$ is obtained by using $\hat{\mathbf{n}}=\hat{\mathbf{z}}$ and the test function $\phi(x,y)=(x/\Delta x)\left((y/\Delta y)^2-1/12\right)$ in eqn. (3.9) to get

$$\frac{1}{2160}\frac{dB^z_{xyy}(t)}{dt} + \frac{1}{\varepsilon}\frac{1}{2\Delta x}\left[\begin{array}{l}\left\langle\left((y/\Delta y)^2 - 1/12\right)D^{y*}(x=\Delta x/2, y)\right\rangle \\ +\left\langle\left((y/\Delta y)^2 - 1/12\right)D^{y*}(x=-\Delta x/2, y)\right\rangle\end{array}\right]$$

$$-\frac{1}{\varepsilon}\frac{1}{6\Delta y}\left[\left\langle(x/\Delta x)D^{x*}(x, y=\Delta y/2)\right\rangle - \left\langle(x/\Delta x)D^{x*}(x, y=-\Delta y/2)\right\rangle\right] \quad \text{(B.13)}$$

$$-\frac{1}{\varepsilon}\frac{1}{\Delta x}\left\{\left((y/\Delta y)^2 - 1/12\right)D^y(x, y)\right\} + \frac{1}{\varepsilon}\frac{2}{\Delta y}\left\{(x/\Delta x)(y/\Delta y)D^x(x, y)\right\} = 0$$

Notice that $D^x(x, y)$ and $D^y(x, y)$ are expressed in terms of an orthogonal basis set in eqns. (B.3) and (B.4). As a result, the integrals associated with the curly brackets in eqns. (B.10), (B.11), (B.12) and (B.13) can be done analytically by using a little bit of algebraic dexterity.

The von Neumann stability analysis at fourth order proceeds analogously to the one in Section III.b for the second order case and Section IV for the third order case. All the eleven variables that were identified in the last paragraph of Section IV as participating in the von Neumann stability analysis at third order also participate in the fourth order case. In addition, we add the variables $D^x_{yyy}(t)$, $D^y_{xxx}(t)$, $a_{xxy}(t)$ and $b_{xyy}(t)$ from our analysis of the fourth order updates of the electric displacement vector field. The vector field for the magnetic induction also adds the four variables $B^z_{xxx}(t)$, $B^z_{yyy}(t)$, $B^z_{xxy}(t)$ and $B^z_{xyy}(t)$ to the von Neumann stability analysis at fourth order. As a result, our amplification matrix will be a $19\times19$ matrix. It is directly obtained by using a computer algebra system.

**Figure Captions**

*Fig. 1 shows us that the primal variables of the DG scheme, given by the normal components and their higher moments for the magnetic induction and electric field displacement. These variables are facially-collocated and are explicitly shown in the figure for the two-dimensional second order accurate DG scheme. They undergo an update from Faraday's law and the generalized Ampere's law respectively. The components of the primal magnetic induction vector and its higher moments are shown by the thick blue arrows while the components of the primal electric displacement vector and its higher moments are shown by the thick red arrows. The edge-collocated electric*

*displacement fields, which are used for updating the facial magnetic induction components, are shown by the thin blue arrows close to the appropriate edge. The edge-collocated magnetic induction fields, which are used for updating the facial electric displacement components, are shown by the thin red arrows close to the appropriate edge.*

*Fig. 2a shows how the facially collocated Fourier modes associated with the electric displacement relate to one another across the different faces of the mesh. Fig. 2b shows the analogous information for the Fourier modes associated with the magnetic induction. These Fourier modes are used in making the von Neumann stability analysis.*

*Fig. 3a shows the amplification factor for wave propagation in various directions relative to the mesh for waves that have a wavelength of five zones. The green, red, cyan and blue curves show the results for the second order P=1 DGTD scheme, the third order P=2 DGTD scheme, the fourth order P=3 DGTD scheme and the Yee scheme respectively. The temporal accuracy matches the spatial accuracy of the scheme. Fig. 3b shows the phase velocity, normalized to unity, for the same four schemes using the same color coding. We see that with increasing order of accuracy, the schemes become closer to the ideal limit. The results correspond to a CFL that is 95% of the maximum.*

*Fig. 4a shows the amplification factor for wave propagation in various directions relative to the mesh for waves that have a wavelength of ten zones. The green, red, cyan and blue curves show the results for the second order P=1 DGTD scheme, the third order P=2 DGTD scheme, the fourth order P=3 DGTD scheme and the Yee scheme respectively. The temporal accuracy matches the spatial accuracy of the scheme. Fig. 4b shows the phase velocity, normalized to unity, for the same four schemes using the same color coding. We see that with increasing order of accuracy, the schemes become closer to the ideal limit. The results correspond to a CFL that is 95% of the maximum.*

*Fig. 5 is analogous to Fig. 3, except that it pertains to the wave propagation at various angles for P0P1, P0P2 , P1P2 and P2P2 schemes for CED. The waves span five zones. The P2P2 scheme is just the P=2 DGTD scheme and is shown for reference. The vertical scales in Fig. 5 are different from Fig. 3. Fig. 5a shows the amplification factor for wave propagation in various directions relative to the mesh for waves that have a wavelength of five zones. The blue, green, red and cyan*

*curves show the results for the second order P0P1 scheme, the third order P0P2 scheme, the third order P1P2 scheme and the third order P2P2 schemes respectively. Fig. 5b shows the phase velocity, normalized to unity, for the same four schemes using the same color coding.*

*Fig. 6 is also analogous to Figs. 5 and 3 because it shows the wave propagation at various angles for fourth order P0P3, P1P3 and P3P3 schemes for CED. The waves span five zones. The P3P3 scheme is just the P=3 DGTD scheme and is shown for reference. The vertical scales in Fig. 6 are different from Figs. 5 and 3. Fig. 6a shows the amplification factor for wave propagation in various directions relative to the mesh for waves that have a wavelength of five zones. The blue, green and red curves show the results for the fourth order P0P3, P1P3 and P3P3 schemes respectively. Fig. 6b shows the phase velocity, normalized to unity, for the same three schemes using the same color coding.*

*Fig. 7 is analogous to Fig. 4, except that it pertains to the wave propagation in various angles for P0P1, P0P2, P1P2 and P2P2 schemes for CED. In this figure the waves span ten zones. The P2P2 scheme is just the P=2 DGTD scheme and is shown for reference. Fig. 7a shows the amplification factor for wave propagation in various directions relative to the mesh for waves that have a wavelength of ten zones. The blue, green, red and cyan curves show the results for the second order P0P1 scheme, the third order P0P2 scheme, the third order P1P2 scheme and the third order P2P2 schemes respectively. Fig. 7b shows the phase velocity, normalized to unity, for the same four schemes using the same color coding.*

*Fig. 8 is also analogous to Figs. 7 and 4 because it shows the wave propagation at various angles for fourth order P0P3, P1P3 and P3P3 schemes for CED. In this figure the waves span ten zones. The P3P3 scheme is just the P=3 DGTD scheme and is shown for reference. The vertical scales in Fig. 8 are different from Figs. 7 and 4. Fig. 8a shows the amplification factor for wave propagation in various directions relative to the mesh for waves that have a wavelength of five zones. The blue, green and red curves show the results for the fourth order P0P3, P1P3 and P3P3 schemes respectively. Fig. 8b shows the phase velocity, normalized to unity, for the same three schemes using the same color coding.*

*Fig 9a shows the electromagnetic energy after one periodic orbit as a function of number of zones along one direction of the two-dimensional mesh for P=1, P=2 and P=3 DG schemes. Fig. 9b*

*shows the same information for P0P1, P0P2 and P0P3 schemes. Fig. 9c shows the same information for P1P2 and P1P3 schemes. All second order schemes are shown in blue; all third order schemes are shown in green; all fourth order schemes are shown in red.*

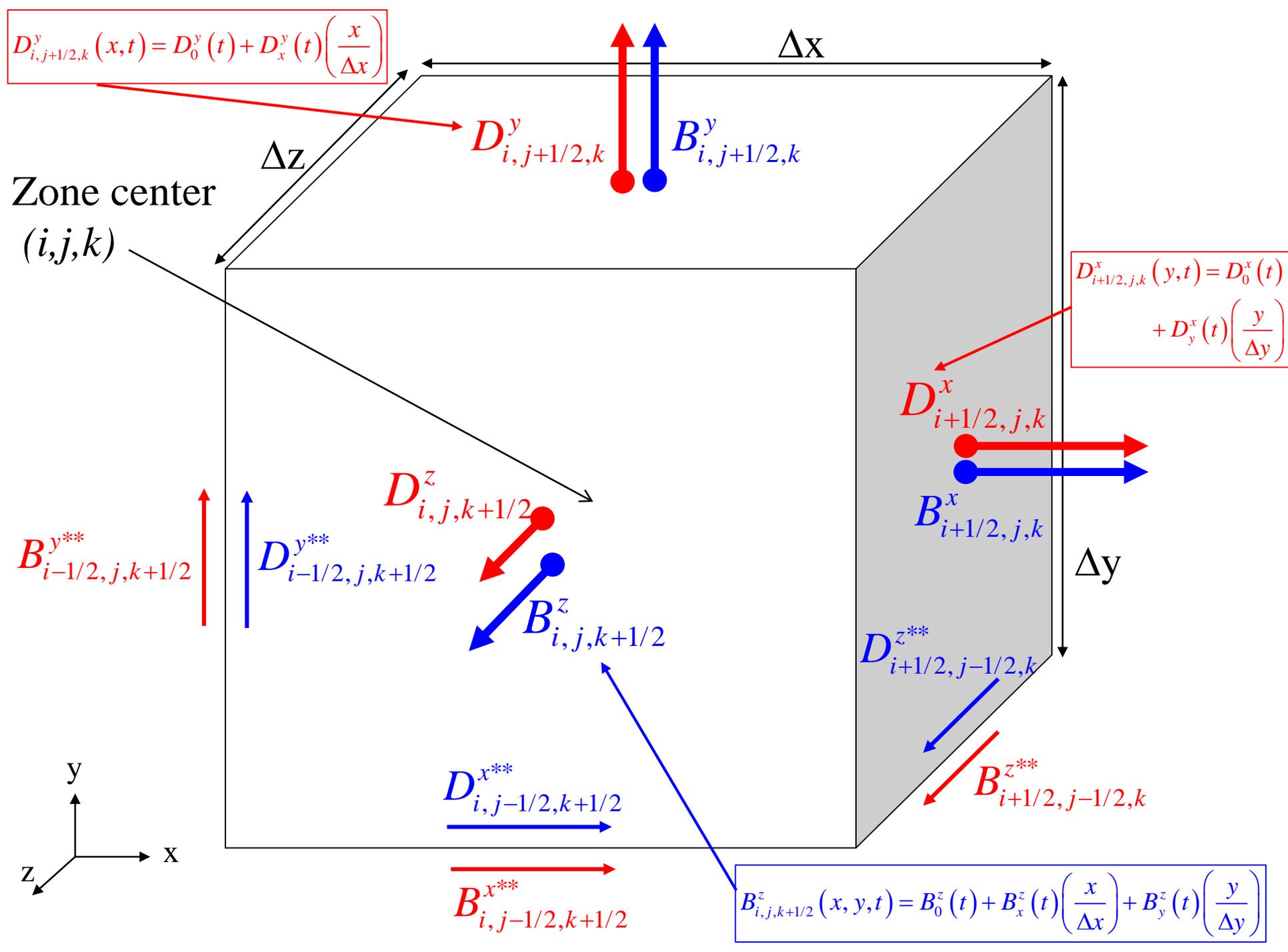

*Fig. 1 shows us that the primal variables of the DG scheme, given by the normal components and their higher moments for the magnetic induction and electric field displacement. These variables are facially-collocated and are explicitly shown in the figure for the two-dimensional second order accurate DG scheme. They undergo an update from Faraday's law and the generalized Ampere's law respectively. The components of the primal magnetic induction vector and its higher moments are shown by the thick blue arrows while the components of the primal electric displacement vector and its higher moments are shown by the thick red arrows. The edge-collocated electric displacement fields, which are used for updating the facial magnetic induction components, are shown by the thin blue arrows close to the appropriate edge. The edge-collocated magnetic induction fields, which are used for updating the facial electric displacement components, are shown by the thin red arrows close to the appropriate edge.*

a)

|  | $D_0^{y+}e^{+ik_y\Delta y}, D_x^{y+}e^{+ik_y\Delta y}$ |  |
|---|---|---|
| (−1,1) | (0,1) | (1,1) |
|  | $D_0^{y+}, D_x^{y+}$ |  |

$D_0^{x+}e^{-2ik_x\Delta x}$  $D_0^{x+}e^{-ik_x\Delta x}$  $D_0^{x+}$  $D_0^{x+}e^{+ik_x\Delta x}$
$D_y^{x+}e^{-2ik_x\Delta x}$  $D_y^{x+}e^{-ik_x\Delta x}$  $D_y^{x+}$  $D_y^{x+}e^{+ik_x\Delta x}$

| (−1,0) | (0,0) | (1,0) |
|---|---|---|

$D_0^{y+}e^{-ik_y\Delta y}, D_x^{y+}e^{-ik_y\Delta y}$

| (−1,−1) | (0,−1) | (1,−1) |
|---|---|---|

$D_0^{y+}e^{-2ik_y\Delta y}, D_x^{y+}e^{-2ik_y\Delta y}$

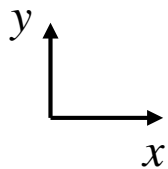

b)

| | $B_0^z e^{+ik_y\Delta y}$ $B_x^z e^{+ik_y\Delta y}, B_y^z e^{+ik_y\Delta y}$ | |
|---|---|---|
| (−1,1) | (0,1) | (1,1) |
| $B_0^z e^{-ik_x\Delta x}$ $B_x^z e^{-ik_x\Delta x}, B_y^z e^{-ik_x\Delta x}$ (−1,0) | $B_0^z$ $B_x^z, B_y^z$ (0,0) | $B_0^z e^{+ik_x\Delta x}$ $B_x^z e^{+ik_x\Delta x}, B_y^z e^{+ik_x\Delta x}$ (1,0) |
| (−1,−1) | $B_0^z e^{-ik_y\Delta y}$ $B_x^z e^{-ik_y\Delta y}, B_y^z e^{-ik_y\Delta y}$ (0,−1) | (1,−1) |

*Fig. 2a shows how the facially collocated Fourier modes associated with the electric displacement relate to one another across the different faces of the mesh. Fig. 2b shows the analogous information for the Fourier modes associated with the magnetic induction. These Fourier modes are used in making the von Neumann stability analysis.*

a) 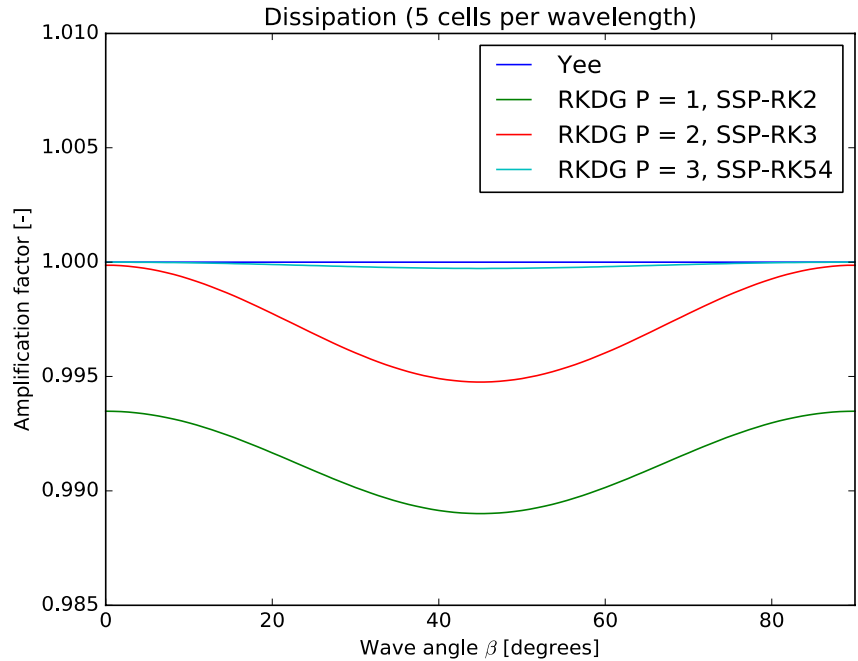 b) 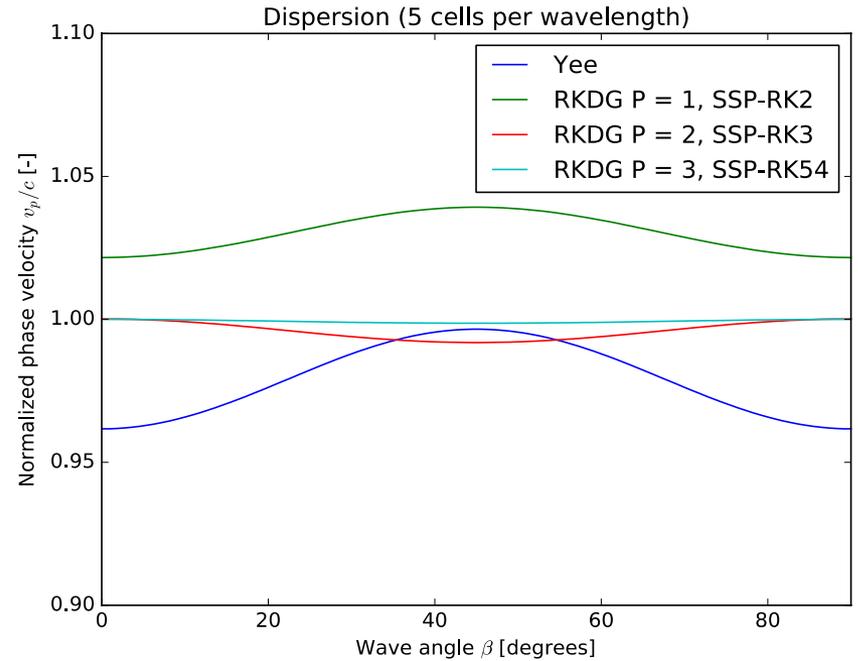

Fig. 3a shows the amplification factor for wave propagation in various directions relative to the mesh for waves that have a wavelength of five zones. The green, red, cyan and blue curves show the results for the second order P=1 DGTD scheme, the third order P=2 DGTD scheme, the fourth order P=3 DGTD scheme and the Yee scheme respectively. The temporal accuracy matches the spatial accuracy of the scheme. Fig. 3b shows the phase velocity, normalized to unity, for the same four schemes using the same color coding. We see that with increasing order of accuracy, the schemes become closer to the ideal limit. The results correspond to a CFL that is 95% of the maximum.

a) 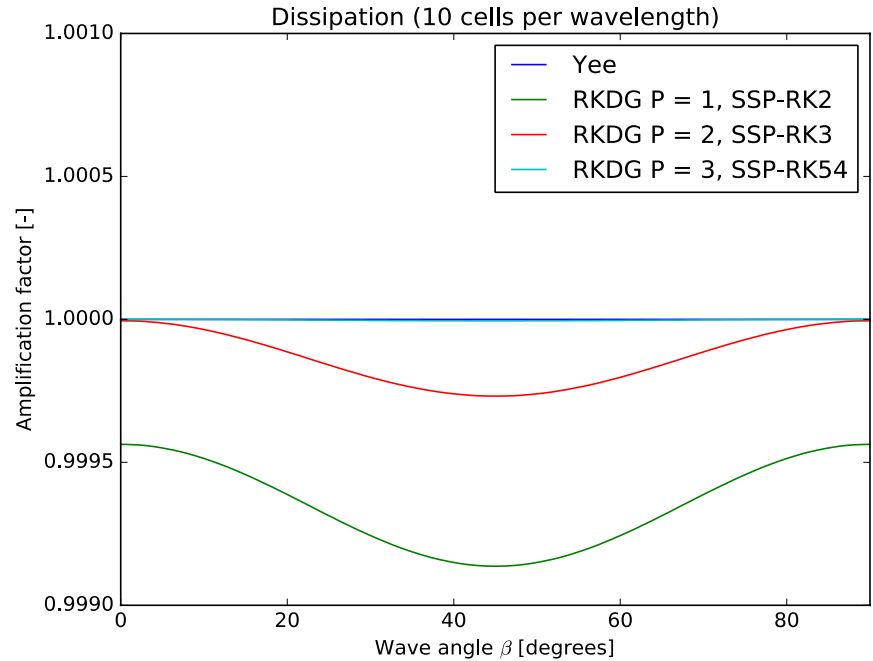 b) 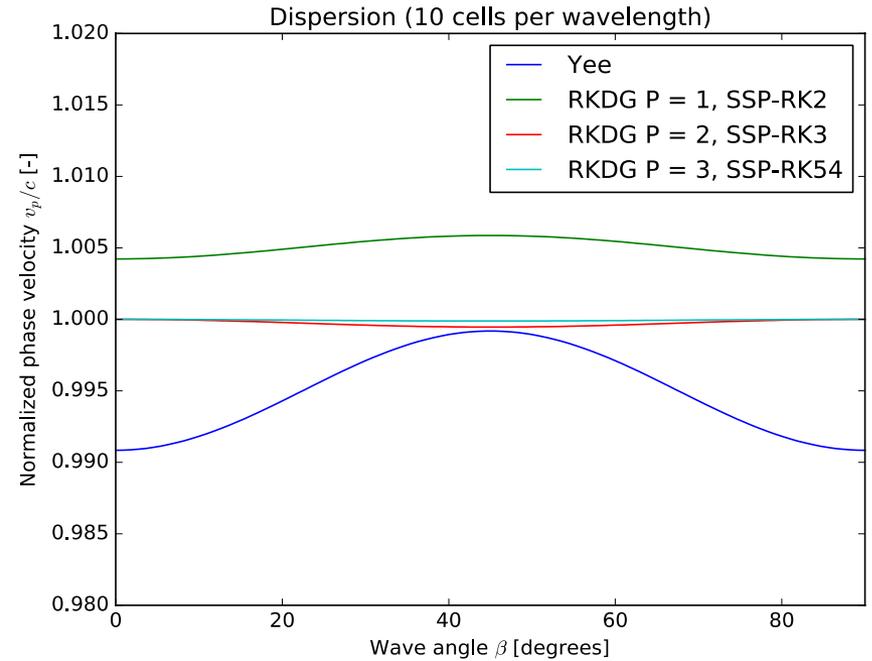

*Fig. 4a shows the amplification factor for wave propagation in various directions relative to the mesh for waves that have a wavelength of ten zones. The green, red, cyan and blue curves show the results for the second order P=1 DGTD scheme, the third order P=2 DGTD scheme, the fourth order P=3 DGTD scheme and the Yee scheme respectively. The temporal accuracy matches the spatial accuracy of the scheme. Fig. 4b shows the phase velocity, normalized to unity, for the same four schemes using the same color coding. We see that with increasing order of accuracy, the schemes become closer to the ideal limit. The results correspond to a CFL that is 95% of the maximum.*

a) 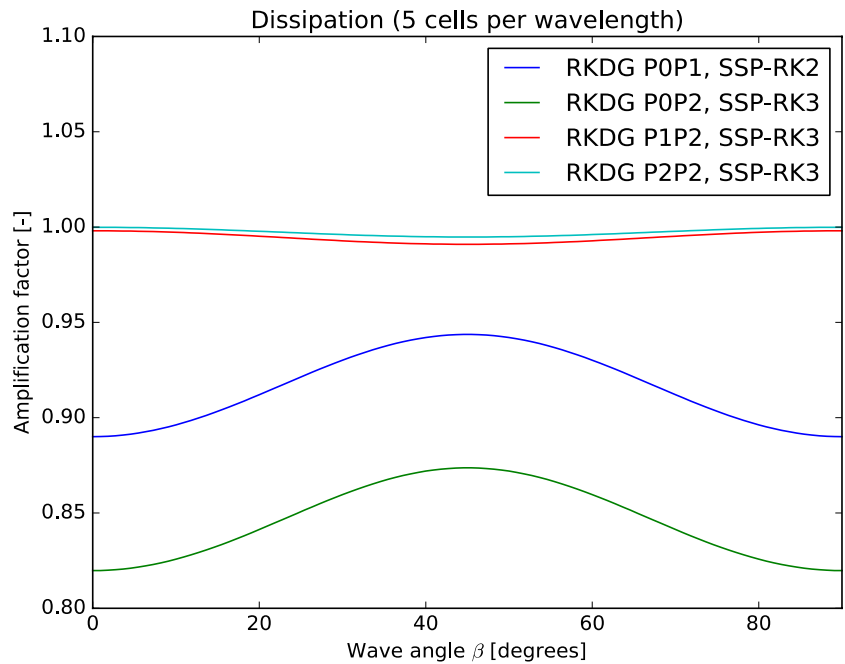 b) 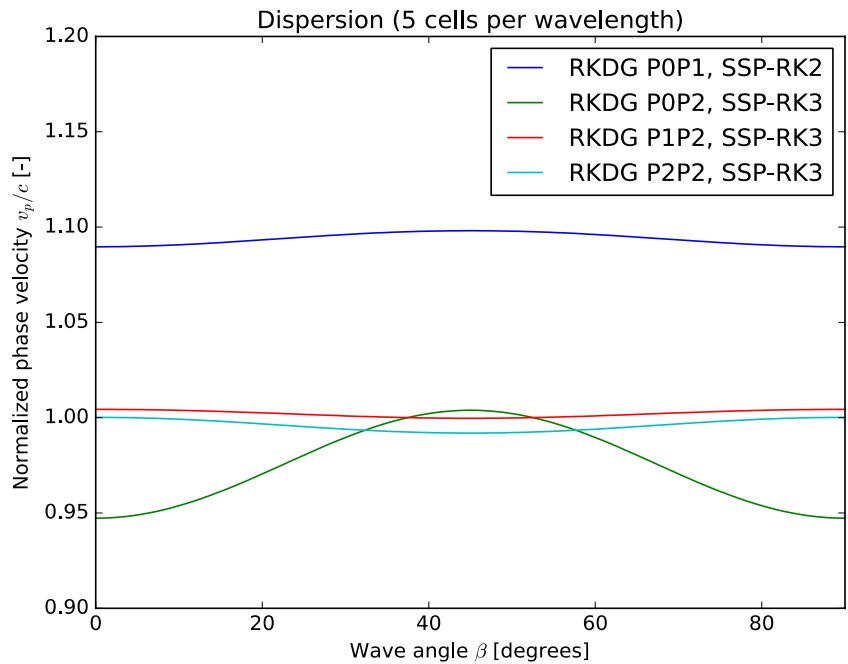

*Fig. 5 is analogous to Fig. 3, except that it pertains to the wave propagation at various angles for P0P1, P0P2, P1P2 and P2P2 schemes for CED. The waves span five zones. The P2P2 scheme is just the P=2 DGTD scheme and is shown for reference. The vertical scales in Fig. 5 are different from Fig. 3. Fig. 5a shows the amplification factor for wave propagation in various directions relative to the mesh for waves that have a wavelength of five zones. The blue, green, red and cyan curves show the results for the second order P0P1 scheme, the third order P0P2 scheme, the third order P1P2 scheme and the third order P2P2 schemes respectively. Fig. 5b shows the phase velocity, normalized to unity, for the same four schemes using the same color coding.*

a) 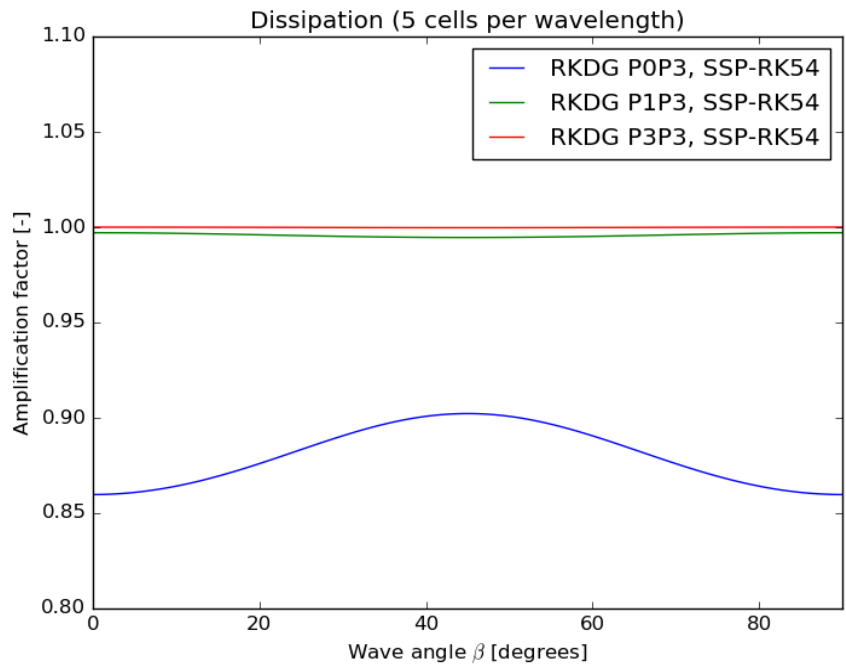
b) 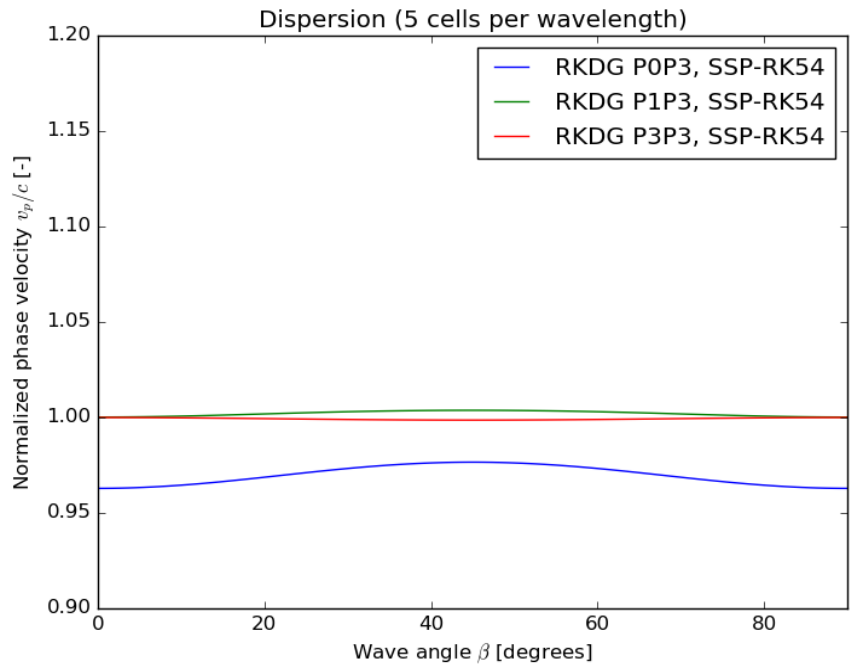

*Fig. 6 is also analogous to Figs. 5 and 3 because it shows the wave propagation at various angles for fourth order P0P3, P1P3 and P3P3 schemes for CED. The waves span five zones. The P3P3 scheme is just the P=3 DGTD scheme and is shown for reference. The vertical scales in Fig. 6 are different from Figs. 5 and 3. Fig. 6a shows the amplification factor for wave propagation in various directions relative to the mesh for waves that have a wavelength of five zones. The blue, green and red curves show the results for the fourth order P0P3, P1P3 and P3P3 schemes respectively. Fig. 6b shows the phase velocity, normalized to unity, for the same three schemes using the same color coding.*

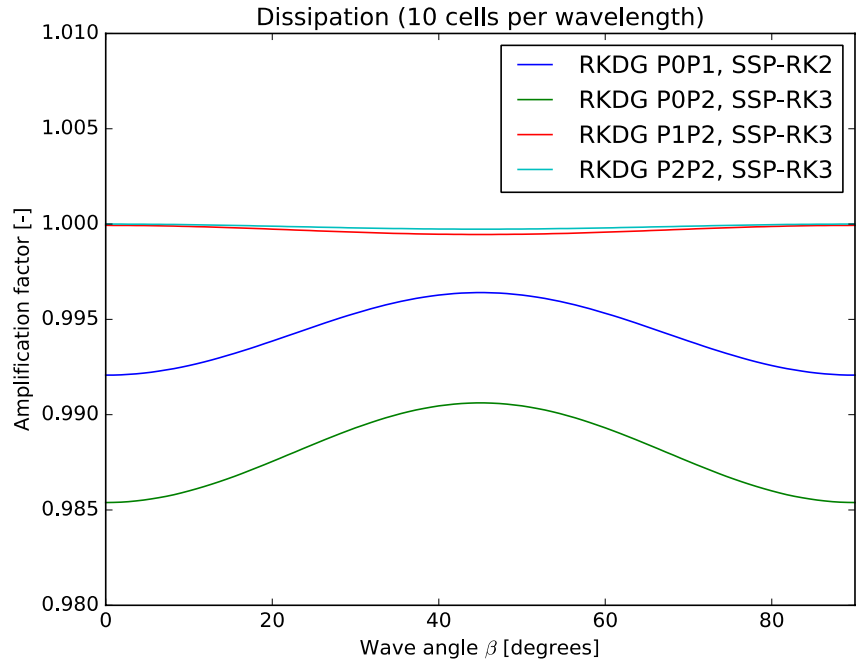 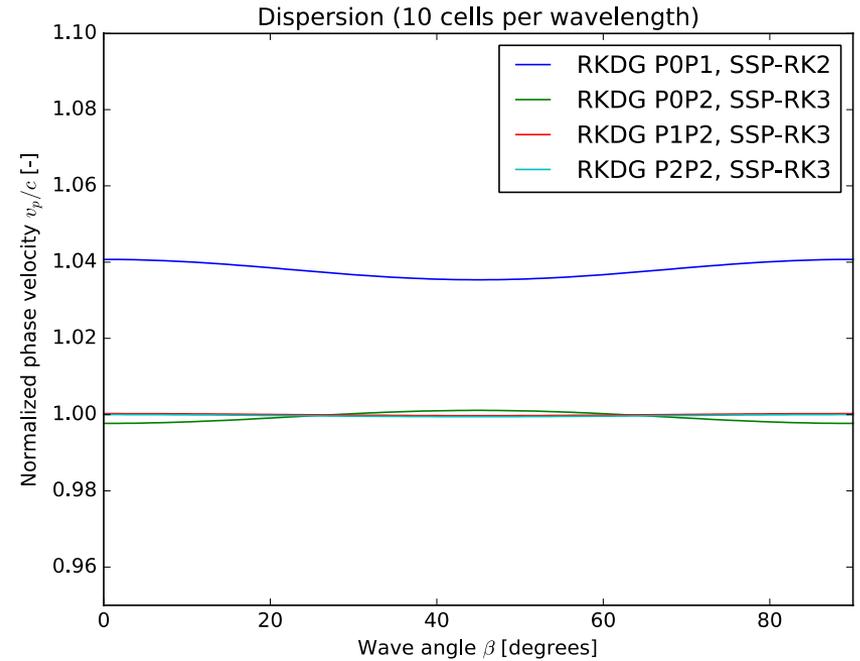

Fig. 7 is analogous to Fig. 4, except that it pertains to the wave propagation in various angles for P0P1, P0P2, P1P2 and P2P2 schemes for CED. In this figure the waves span ten zones. The P2P2 scheme is just the P=2 DGTD scheme and is shown for reference. Fig. 7a shows the amplification factor for wave propagation in various directions relative to the mesh for waves that have a wavelength of ten zones. The blue, green, red and cyan curves show the results for the second order P0P1 scheme, the third order P0P2 scheme, the third order P1P2 scheme and the third order P2P2 schemes respectively. Fig. 7b shows the phase velocity, normalized to unity, for the same four schemes using the same color coding.

a) 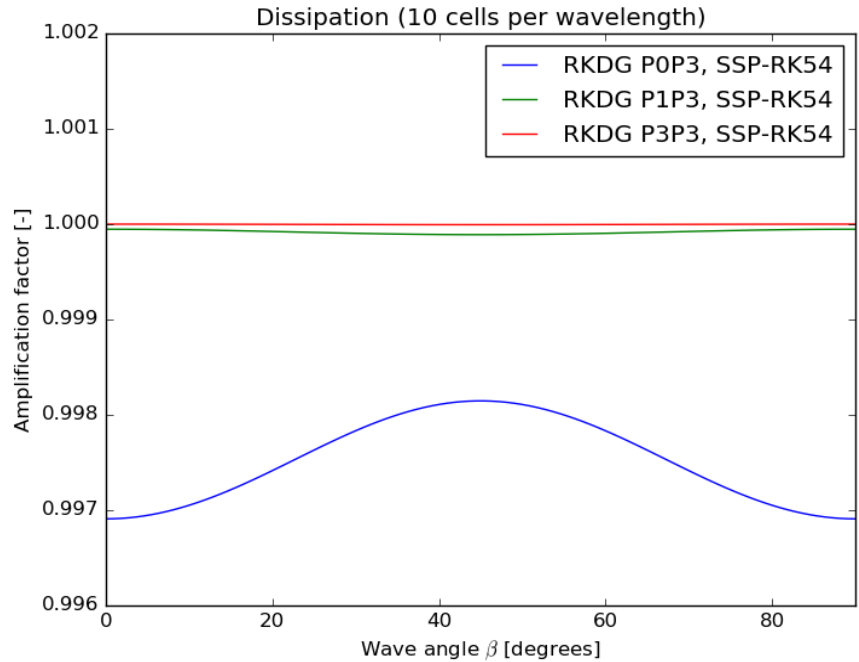
b) 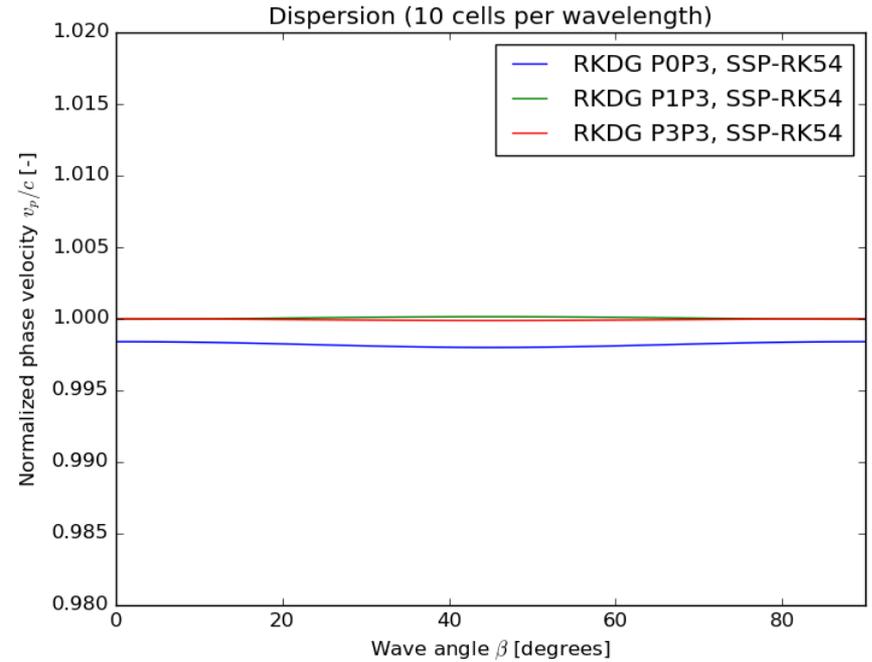

*Fig. 8 is also analogous to Figs. 7 and 4 because it shows the wave propagation at various angles for fourth order P0P3, P1P3 and P3P3 schemes for CED. In this figure the waves span ten zones. The P3P3 scheme is just the P=3 DGTD scheme and is shown for reference. The vertical scales in Fig. 8 are different from Figs. 7 and 4. Fig. 8a shows the amplification factor for wave propagation in various directions relative to the mesh for waves that have a wavelength of five zones. The blue, green and red curves show the results for the fourth order P0P3, P1P3 and P3P3 schemes respectively. Fig. 8b shows the phase velocity, normalized to unity, for the same three schemes using the same color coding.*

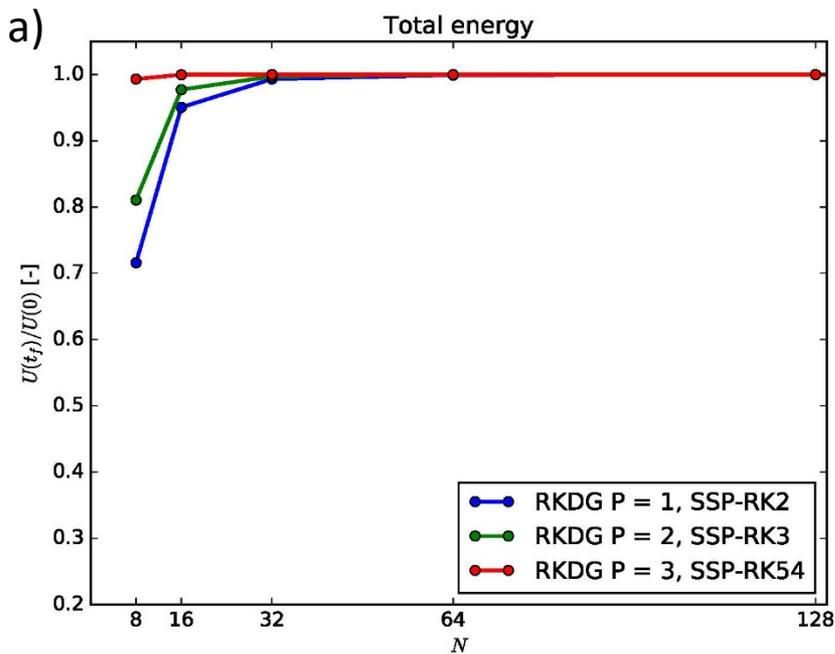

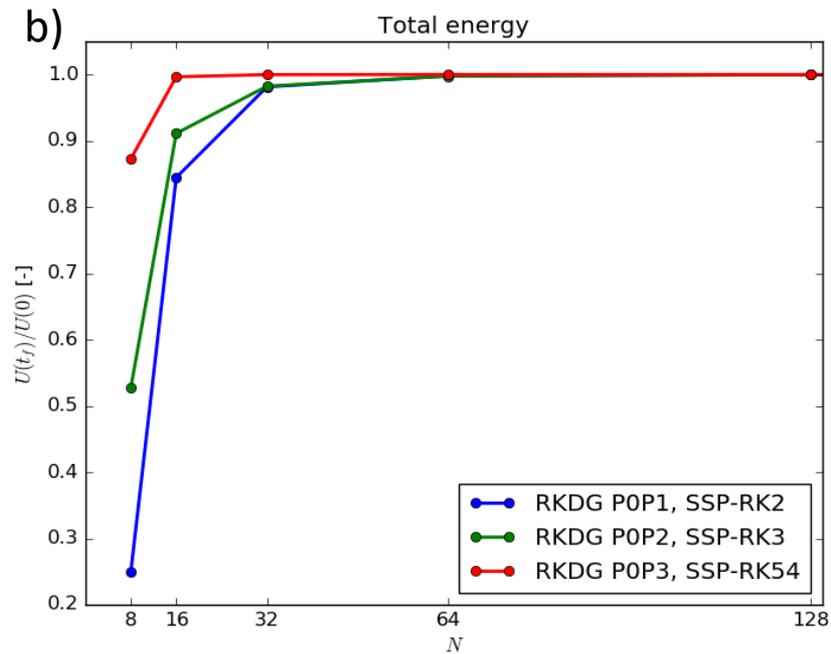

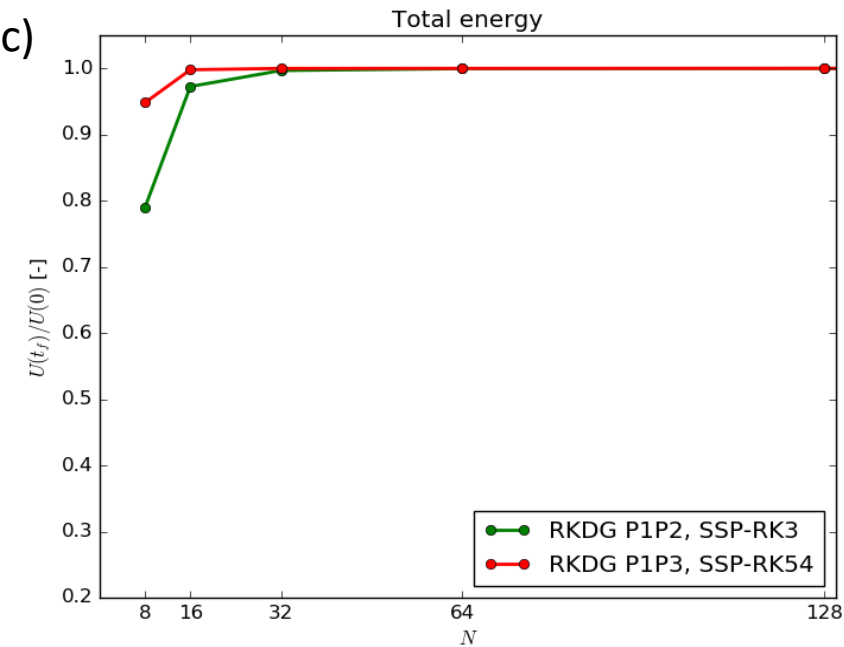

*Fig 9a shows the electromagnetic energy after one periodic orbit as a function of number of zones along one direction of the two-dimensional mesh for P=1, P=2 and P=3 DG schemes. Fig. 9b shows the same information for P0P1, P0P2 and P0P3 schemes. Fig. 9c shows the same information for P1P2 and P1P3 schemes. All second order schemes are shown in blue; all third order schemes are shown in green; all fourth order schemes are shown in red.*